\documentclass[preprint,12pt]{elsarticle}

\usepackage{amsmath}
\usepackage{amssymb}
\usepackage{amsfonts}
\usepackage{newtxtext} % Times New Roman-like text font
\usepackage{newtxmath} % Times New Roman-like math font

\usepackage[table,dvipsnames]{xcolor}
\usepackage{graphicx}
\usepackage{float}

\usepackage{array}     % Enhanced column formatting
\usepackage{multirow}  % Multi-row cells
\usepackage{makecell}  % Line breaks in table cells
\usepackage{tabularx}  % Auto-width tables
\usepackage{booktabs}  % Professional table rules
\usepackage{arydshln}  % Dashed lines in tables; load after array/booktabs/tabularx

\usepackage[ruled,vlined,linesnumbered]{algorithm2e}
\usepackage{setspace}       % Provides \setstretch
\usepackage[most]{tcolorbox} % Provides tcolorbox and enhanced options

\usepackage[unicode]{hyperref}
\newcolumntype{Y}{>{\centering\arraybackslash}X}

\journal{Expert Systems With Applications}

\hypersetup{
  pdfauthor={Yixuan Li, Mingxuan Huang, Jiajing Wang, Weidong Yang, Xinyi Liu, Ben Fei, Lipeng Ma},
  pdftitle={SpecCoder},
  pdfkeywords={Code Generation, Specification-Aware Reasoning, Reinforcement Learning, Large Language Models}
}

\begin{document}

\begin{frontmatter}

\title{SpecCoder: Specification-Aware Code Generation with Curriculum Dual-Task Reinforcement Learning}

\author[fudan]{Yixuan Li}
\ead{yxli24@m.fudan.edu.cn}

\author[fudan]{Mingxuan Huang}
\ead{24307130121@m.fudan.edu.cn}

\author[fudan]{Jiajing Wang}
\ead{25213050375@m.fudan.edu.cn}

\author[fudan]{Weidong Yang\corref{cor1}}
\ead{wdyang@fudan.edu.cn}

\author[fudan]{Xinyi Liu}
\ead{liuxiny24@m.fudan.edu.cn}

\author[cuhk]{Ben Fei}
\ead{benfei@cuhk.edu.hk}

\author[fudan]{Lipeng Ma\corref{cor1}}
\ead{lpma21@m.fudan.edu.cn}

\cortext[cor1]{Corresponding authors}

\address[fudan]{School of Computer Science, Fudan University, Shanghai, 200438, China}
\address[cuhk]{Department of Information Engineering, The Chinese University of Hong Kong, Hong Kong, 999077, China}

%% Abstract
\begin{abstract}

Large language models (LLMs) have made substantial progress in code generation but still struggle with challenging programming tasks that require understanding rich natural language requirements. 
These requirements often specify problem goals, input/output formats, constraints, examples, edge cases, and implicit assumptions.
Overlooking even 
one of them may produce executable but functionally incorrect code.
\textcolor{black}{Existing 
training-free methods mainly rely on prompting or agent-based workflows, while 
training-based methods typically optimize final code outputs using supervised or 
execution-based signals. However, existing approaches provide limited supervision for learning the intermediate mapping from raw requirements to structured specification analyses and for grounding such analyses in concrete implementation behavior. 
% Consequently, models may overlook critical constraints or produce implementations that deviate from their own specification analyses.
Consequently, models may omit critical constraints when interpreting raw requirements, and even when an explicit specification analysis is produced, the resulting implementation may fail to reflect it consistently.
}
Motivated by this gap, we propose SpecCoder, a specification-aware two-stage training framework for code generation. SpecCoder first employs specification-guided Supervised Fine-Tuning (SFT) to train LLMs to derive structured specification analyses and generate code conditioned on them. 
It then introduces curriculum dual-task Group Relative Policy Optimization (GRPO), which jointly optimizes specification-guided generation and discrimination to \textcolor{black}{encourage stronger correspondence} between structured specification analyses and code behavior. 
Experiments on APPS, CodeContests, and xCodeEval demonstrate the effectiveness of specification-aware training, with SpecCoder consistently improving both standalone code generation and the performance of agent-based workflows.
\textcolor{black}{Additional evaluations on BigCodeBench-Hard and ClassEval, together with human evaluation and specification perturbation studies, further support the effectiveness of structured specification analyses and their role in guiding code generation and discrimination.}
\end{abstract}

%% Keywords

\end{frontmatter}

\section{Introduction}
\label{label:section_intro}

Code generation is a fundamental research problem in software engineering.
Large language models (LLMs) have made substantial progress and perform well on widely used benchmarks such as HumanEval~\cite{chen2021evaluating} and MBPP~\cite{austin2021program}.
However, their performance remains limited on more challenging programming benchmarks, such as APPS~\cite{hendrycks2021measuring}, CodeContests~\cite{li2022competition}, and xCodeEval~\cite{khan2024xcodeeval}.
One key difference is that benchmarks such as HumanEval and MBPP usually provide relatively simple prompts, such as a function signature or a short instruction, whereas more challenging tasks often require LLMs to reason over richer raw natural language requirements.
These requirements specify the intended code behavior and may include problem goals, input and output formats, constraints, examples, edge cases, and implicit assumptions. 
Therefore, solving such specification-rich tasks requires LLMs not only to generate syntactically valid or executable code, but also to form an explicit understanding of the requirements and ensure that the implementation satisfies the intended functionality and constraints.

Existing methods for improving LLM-based code generation can be broadly categorized into \textbf{training-free} and \textbf{training-based} approaches.
\textbf{Training-free} methods can be further divided into prompt-based methods and agent-based workflows.
Prompt-based methods~\cite{olausson2023selfrepair,zhang2023self,jiang2024selfplanning,dong2024selfcollaboration} guide the generation process through carefully designed instructions for planning, reasoning, or output repair.
They are lightweight and easy to deploy, but their effectiveness is often limited because requirement understanding remains implicit and is not directly optimized.
Agent-based methods~\cite{hong2023metagpt,huang2023agentcoder,mathews2024tgen,zhang2024paircoder} decompose code generation into multi-step or multi-role workflows involving requirement analyses, implementation, testing, and revision.
Although effective, these methods typically require repeated interactions and iterative execution, leading to substantial inference cost.
More importantly, both prompt-based and agent-based methods mainly improve how LLMs use prompts or external workflows at inference time, rather than enhancing their intrinsic ability to align generated code with raw requirements.
\textbf{Training-based} methods~\cite{zhang2025codedpo,wang2025rlcoder,li2025codeprm,jiang2025coderl+} optimize LLMs through reinforcement learning or preference optimization, primarily using execution-based feedback from test cases or preference signals.
However, such feedback often provides sparse reward signals and does not explicitly supervise how requirements are understood and translated into structured specification analyses.
\textcolor{black}{Consequently, existing training-based methods provide limited supervision for learning to derive structured specification analyses from raw requirements and use them to guide code generation.}

Recent studies~\cite{tian2025aligning,li2026bridging} show that code generation failures often stem from misunderstanding raw requirements.
When requirements such as key concepts, input and output constraints, or edge cases are overlooked, generated code can easily deviate from the intended code behavior.
However, existing specification-alignment methods~\cite{chen2025revisit,tian2025ufiX} mainly perform alignment, repair, or verification during inference, rather than improving the intrinsic specification-alignment ability of LLMs through training.
A natural direction is therefore to train LLMs with explicit specification-aware reasoning capabilities. This direction presents two challenges.
First, building an accurate and structured specification-level understanding from raw requirements remains challenging.
For complex programming tasks, requirement understanding should be made explicit before implementation, as implicit contextual reasoning alone may cause LLMs to omit, misinterpret, or weaken key constraints.
Second, generated code is not always well aligned with the specification-level understanding.
Even when reasonable structured specification analyses are produced, the final implementation may still ignore, simplify, or contradict them, creating a gap between the stated understanding and actual code behavior.
Moreover, execution-based feedback typically provides only outcome-level signals~\cite{le2022coderl,liu2023rltf,li2025codeprm}, indicating whether code passes tests but not which requirement constraints are satisfied or violated.
As a result, such feedback provides limited guidance for training models to follow structured specifications or compare candidate implementations according to specification satisfaction.

To address these challenges, we propose \textbf{SpecCoder}, a specification-aware two-stage training framework for code generation.
SpecCoder first conducts specification-guided Supervised Fine-Tuning (SFT), enabling the LLM to derive structured specification analyses from raw requirements and generate code conditioned on them.
SpecCoder then applies curriculum dual-task Group Relative Policy Optimization (GRPO).
In this stage, the generation task further optimizes specification-guided code generation, while the discrimination task trains the LLM to select, from paired candidate implementations, the one that better satisfies the raw requirement under the shared structured specification analysis.
For both GRPO tasks, SpecCoder categorizes training samples into Easy, Medium, and Hard subsets and organizes them into a three-phase curriculum.
The curriculum first establishes basic associations between structured specification analyses and code behavior using easier samples, then addresses harder cases requiring complex requirement analysis and implementation comparison, and finally mixes all difficulty levels to consolidate learned behaviors and mitigate forgetting.
By combining these complementary generation and discrimination objectives, SpecCoder better aligns structured specification analyses with code behavior.

We evaluate SpecCoder as a standalone code generator and as the backbone of representative agent-based code generation frameworks.
In the standalone setting, SpecCoder achieves strong performance on the in-distribution (ID) APPS and CodeContests benchmarks, reaching Pass@1/AvgPassRatio scores of 0.186/0.346 and 0.091/0.207, respectively.
It achieves the best results among training-free prompting baselines and is competitive with or superior to training-based baselines across almost all metrics.
On the out-of-distribution (OOD) xCodeEval benchmark, SpecCoder obtains a Pass@1 of 0.147 and an AvgPassRatio of 0.324, outperforming all compared training-free prompting and training-based baselines.
When used as the backbone of agent-based frameworks, SpecCoder consistently improves all evaluated frameworks across APPS, CodeContests, and xCodeEval, with gains of up to 11.3 percentage points in Pass@1 and 13.2 percentage points in AvgPassRatio over the base model.
\textcolor{black}{These results demonstrate that SpecCoder improves both standalone code generation and its use as a backbone for agent-based workflows. 
Further analyses show that structured specification analyses provide useful intermediate guidance for code synthesis.}

This paper makes the following contributions.

\textcolor{black}{
\begin{itemize}
\item We propose \textbf{SpecCoder}, a specification-aware two-stage training framework that combines specification-guided SFT with curriculum dual-task GRPO to explicitly learn structured specification analyses from raw requirements and use them to guide code generation.
\item We introduce structured specification analyses as a shared intermediate representation for specification-guided generation and implementation discrimination, allowing the two training objectives to jointly relate requirement understanding to concrete code behavior.
\item We conduct extensive evaluations of SpecCoder in standalone and agent-based settings across ID and OOD benchmarks. Additional analyses on intermediate representations, specification dependency, and realistic programming tasks further demonstrate its effectiveness and applicability.
\end{itemize}
}

\section{Related Work}

\subsection{Training-Free Methods for LLM-Based Code Generation}

Training-free methods improve LLM-based code generation without updating model parameters and can be broadly divided into prompt-based methods and agent-based workflows.
Prompt-based methods mainly enhance code generation through structured reasoning and iterative refinement during inference.
Chain-of-Thought prompting~\cite{wei2022chain} shows that intermediate reasoning can improve LLM performance on complex tasks, motivating structured reasoning approaches for code generation.
Self-Debug~\cite{chen2024teaching,chen2025revisit} and Self-Edit~\cite{zhang2023self} iteratively repair generated code based on execution feedback, whereas later methods including Self-Planning~\cite{jiang2024selfplanning}, SCoT~\cite{li2025SCoT}, Program of Thoughts~\cite{chen2022program}, \textcolor{black}{ArchCode}~\cite{han2024archcode}, and $\mu$FiX~\cite{tian2025ufiX} \textcolor{black}{introduce structured pre-generation reasoning, explicitly incorporate software requirements, or integrate specification understanding with execution-based refinement.}
Despite their training-free nature, these methods remain limited because manually designed prompts are often less robust, and LLMs may struggle to consistently follow complex instructions or retain critical requirement details.

Agent-based methods further improve complex code generation by decomposing the development process into multi-step or multi-role workflows.
Frameworks such as MetaGPT~\cite{hong2023metagpt} and FlowGen~\cite{lin2025soen} simulate software development procedures through cooperative agents, while Reflexion~\cite{shinn2023reflexion} refines model behavior through verbal feedback.
Recent systems such as Specine~\cite{tian2025aligning} and REA-Coder~\cite{li2026bridging} further address requirement misunderstanding and specification alignment during inference.
These methods can better handle complex requirements by introducing explicit analysis, feedback, or revision steps, but they usually require repeated interactions, tool calls, or iterative execution, leading to higher inference cost.
\textcolor{black}{Overall, these prompt-based and agent-based methods treat requirement analysis 
as an auxiliary inference-time procedure, without directly optimizing either the 
completeness of the derived specification or its consistency with the generated 
implementation. Consequently, a model may produce a plausible requirement analysis 
while still overlooking critical constraints or generating code that contradicts 
that analysis.}

\subsection{Training-Based Methods for LLM-Based Code Generation}

\textcolor{black}{Recent training-based methods for LLM-based code generation can be broadly categorized into feedback-based optimization and training-strategy optimization.
Feedback-driven approaches like CodeRL~\cite{le2022coderl}, CodeRL+~\cite{jiang2025coderl+}, CodeDPO~\cite{zhang2025codedpo}, and CodePRM~\cite{li2025codeprm} enhance code correctness using execution feedback, preference signals, or execution-derived process supervision.
Meanwhile, training-strategy optimization methods improve learning efficiency and exploration through progressive task organization and capability decomposition. 
For instance, StepCoder~\cite{dou2024stepcoder} incrementally increases code-completion difficulty to optimize executed segments. 
RECRL~\cite{yin2026recrl} integrates test-driven difficulty estimation with adaptive curriculum sampling and requirement rewriting. 
Similarly, FixAudit~\cite{tang2026fixaudit} sequentially trains models on execution reasoning, code repair, and defect-test generation. 
These strategies improve exploration, training efficiency, or iterative code refinement.}

\textcolor{black}{Despite these advancements, existing learning signals mainly focus on final implementation outcomes, with limited supervision for accurately understanding raw requirements. 
Consequently, models may omit essential constraints during requirement interpretation, and their final code may fail to consistently reflect the generated intermediate analyses. 
SpecCoder explicitly addresses these two interconnected gaps through a two-stage framework.
It first uses specification-guided SFT to learn structured specification analyses and specification-conditioned generation.
Curriculum dual-task GRPO then jointly optimizes generation and discrimination to strengthen the correspondence between requirement analysis and code behavior.
}

\section{Problem Formulation}
\label{label_section_formulation}
SpecCoder trains LLMs to derive structured specification analyses from raw requirements and generate code aligned with these analyses.
We formulate this objective as two learning tasks: specification-guided code generation and specification-guided code discrimination.
Let $\mathcal{P}$, $\mathcal{S}$, and $\mathcal{C}$ denote the spaces of raw requirements, structured specification analyses, and candidate implementations, respectively.
For each raw requirement $p \in \mathcal{P}$, let $\mathcal{T}_p$ be its test suite.
Given a candidate implementation $c \in \mathcal{C}$, we define $f(c,\mathcal{T}_p) \in [0,1]$ as the fraction of test cases passed by $c$, where $f(c,\mathcal{T}_p)=1$ means that $c$ passes all available tests.
We use $\mathcal{T}_p$ as an executable proxy for observable functional correctness under available tests. Since available tests may not cover all semantic constraints, $f(c,\mathcal{T}_p)$ is treated as an approximate training and evaluation signal rather than a complete measure of requirement satisfaction.

\subsection{Specification-Guided Code Generation Task}

Directly generating code from raw requirements is challenging because LLMs may overlook key goals, constraints, edge cases, or implicit assumptions.
To make requirement analysis explicit, SpecCoder generates a structured output in which a structured specification analysis $\hat{s}$ is produced before the code $\hat{c}$, so that the implementation is guided by an explicit intermediate analysis of the requirement.
Formally, let $\pi_{\theta}$ denote the LLM policy parameterized by $\theta$.
The generation process is defined as:
\begin{equation}
(\hat{s}, \hat{c}) \sim \pi_{\theta}(\cdot \mid p),
\end{equation}
where $\hat{s}$ serves as an intermediate specification-level representation rather than a unique ground-truth analysis.
The output is structured so that $\hat{s}$ is generated before $\hat{c}$, making code generation conditioned on the raw requirement $p$ and the preceding specification analysis $\hat{s}$ in the autoregressive context.

\subsection{Specification-Guided Code Discrimination Task}
To strengthen the alignment between structured specification analyses and code behavior, we introduce a specification-guided discrimination task in which the LLM compares two candidate implementations under the same raw requirement and structured specification analysis.
For each pair of candidate implementations, we construct $(c^+, c^-)$ such that:
\begin{equation}
f(c^+,\mathcal{T}_p) > f(c^-,\mathcal{T}_p),
\end{equation}
where $c^+$ achieves a higher test pass rate than $c^-$, indicating that it better satisfies the requirements under the available tests.
To reduce position bias, the two candidates are randomly ordered as $(c_A, c_B)$, and the LLM analyses them with respect to the shared specification to output a binary decision $y \in \{A,B\}$.

\section{Methodology}

\begin{figure}[htbp]
\centering
\includegraphics[width=\linewidth]{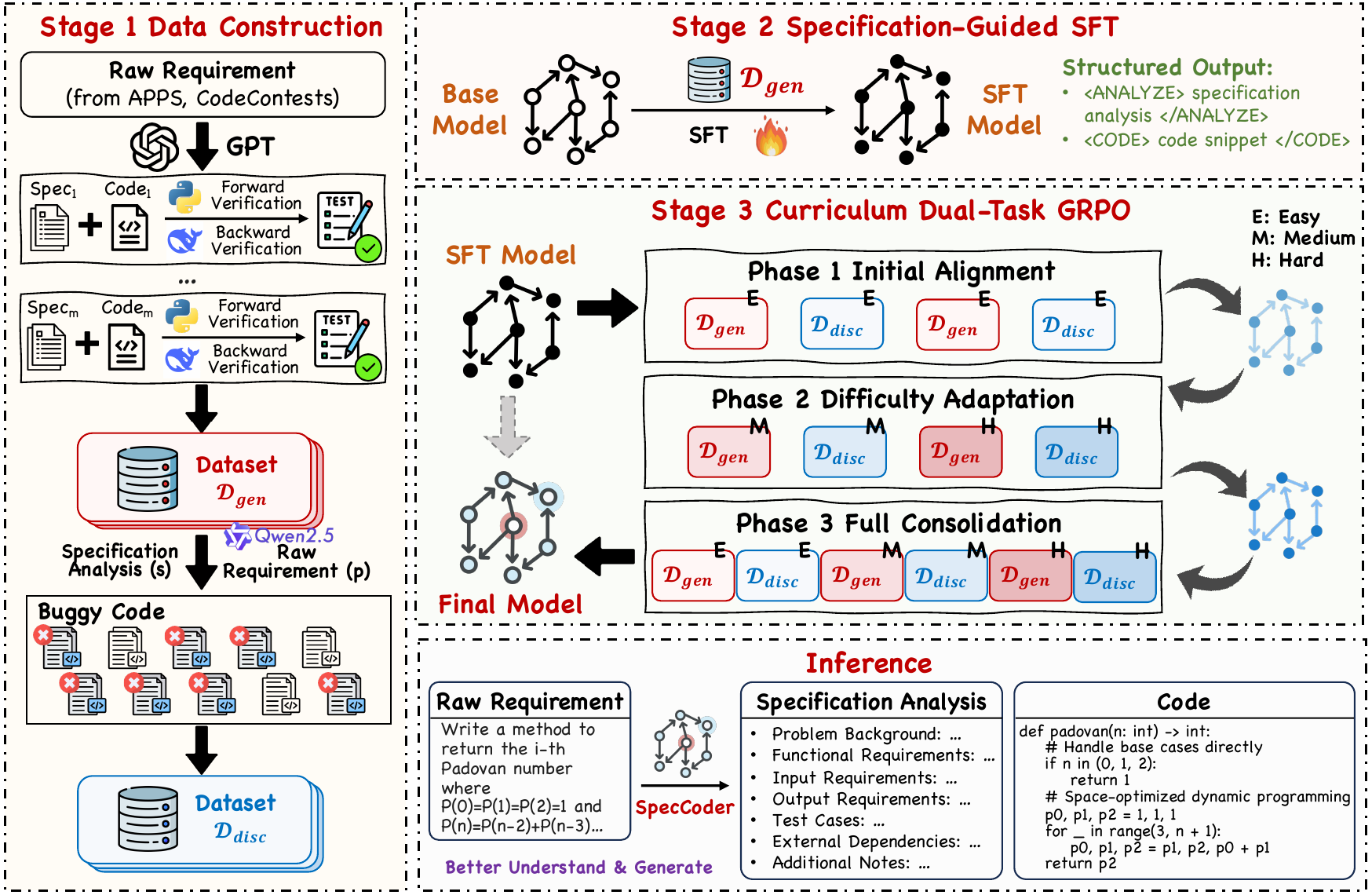}
    \caption{Overview of the SpecCoder pipeline. The methodology consists of a data construction stage followed by two training stages. Stage 1 constructs the specification-guided generation dataset $\mathcal{D}_{\text{gen}}$ through forward execution-based verification and backward semantic verification, and builds the specification-guided discrimination dataset $\mathcal{D}_{\text{disc}}$ from paired candidate implementations under shared problems and specification analyses. Stage 2 performs supervised fine-tuning on $\mathcal{D}_{\text{gen}}$ to teach the LLM to derive structured specification analyses and generate code conditioned on them. Stage 3 applies curriculum dual-task GRPO to jointly optimize specification-guided generation and discrimination, progressively moving from easier samples to harder cases and then to a full mixture for consolidation. During inference, the trained model first analyses the raw requirement into a structured specification analysis and then uses it to guide final code generation.}
\label{figs:overview}
\end{figure}

Figure~\ref{figs:overview} illustrates the overall pipeline of SpecCoder, which contains one data construction stage and two training stages.
SpecCoder uses structured specification analyses as an intermediate interface between raw requirements and candidate implementations.
In the data construction stage, we build a specification-guided generation dataset $\mathcal{D}_{\text{gen}}$ with tuples $(p,s,c)$ and a specification-guided discrimination dataset $\mathcal{D}_{\text{disc}}$ with tuples $(p,s,c_A,c_B,y)$, where $y$ indicates the candidate implementation with the higher test pass rate under the shared raw requirement and structured specification analysis.
In the first training stage, SpecCoder performs specification-guided SFT on $\mathcal{D}_{\text{gen}}$ to initialize the LLM's ability to derive structured specification analyses and generate code conditioned on them.
In the second training stage, SpecCoder applies curriculum dual-task GRPO to jointly optimize specification-guided generation and discrimination, further aligning specification-level understanding with code behavior.
The following sections describe data construction in Section~\ref{sec:data_construction}, specification-guided SFT in Section~\ref{sec:sft}, and curriculum dual-task GRPO in Section~\ref{sec:grpo}.

\subsection{Data Construction Pipeline}
\label{sec:data_construction}
As shown in Stage 1 of Figure~\ref{figs:overview}, the data construction pipeline builds two datasets, namely the specification-guided code generation dataset $\mathcal{D}_{\text{gen}}$ and the specification-guided code discrimination dataset $\mathcal{D}_{\text{disc}}$.
Both datasets are constructed from APPS and CodeContests.
For APPS, we use the original training set for data construction and further split the original test set into two non-overlapping subsets: one used only for additional data construction and the other reserved for evaluation, thereby preventing overlap between construction and evaluation.
In contrast, CodeContests uses only its complete training set.

\subsubsection{Specification-Guided Code Generation Dataset \texorpdfstring{$\mathcal{D}_{\text{gen}}$}{D\_gen}}

The dataset $\mathcal{D}_{\text{gen}}$ provides verified triplets for specification-guided code generation training and is constructed through multi-path generation and bidirectional verification.

\textbf{Multi-Path Generation.} 
For each raw requirement $p$, we use \textbf{\textit{GPT-4o}} (\textit{gpt-4o-2024-08-06}) with a sampling temperature of $\tau=0.7$.
By independently querying the model $M$ times, we generate a set of structured specification analyses ${s^{(1)}, s^{(2)}, \dots, s^{(M)}}$, where $M=3$ in our implementation.
Each specification analysis follows a seven-dimensional structure consisting of problem background, functional requirements, input requirements, output requirements, test cases, external dependencies, and additional notes, as summarized in Table~\ref{tab:spec_dimensions}.
The structure is adapted from common software requirement specification practices~\cite{iso29148,ieee830} and tailored to programming tasks, enabling raw requirements to be organized into explicit specification-level analyses for code generation.
For each generated structured specification analysis $s^{(m)}$, \textbf{\textit{GPT-4o}} further generates a corresponding candidate implementation $c^{(m)}$ conditioned on both $p$ and $s^{(m)}$, forming a candidate triplet $\langle p, s^{(m)}, c^{(m)} \rangle$.

\textbf{Bidirectional Verification.} 
For the candidate triplets generated above, we apply two verification steps to ensure both executable correctness and specification consistency.
Forward verification evaluates functional correctness by executing each candidate implementation against the available test suite $\mathcal{T}_p$.
Only triplets whose candidate implementation satisfies $f(c^{(m)}, \mathcal{T}_p)=1$ are retained.
Backward verification further examines whether the structured specification analysis is consistent with the raw requirement and whether the generated code follows the specification analysis.
Although a candidate implementation may pass all available tests, it may still only partially reflect the specification analysis or rely on assumptions not explicitly captured by it.
To reduce this risk, we use \textbf{\textit{DeepSeek-V3-0324}}~\cite{deepseek2025v30324} as an evaluator to assess each forward-verified triplet $\langle p, s^{(m)}, c^{(m)} \rangle$. 
For space considerations, we evaluate each triplet from three aspects: semantic completeness, algorithmic traceability, and expression clarity, with detailed criteria provided in ~\ref{appendix_verification}.
Triplets satisfying the backward verification criteria are retained. 
If multiple verified triplets remain for a problem, we randomly select one, while problems without any verified triplet are entirely discarded. 
The resulting triplets form the final specification-guided code generation dataset $\mathcal{D}_{\text{gen}}$.

\begin{table}[htbp]
    \centering
    \small
    \setlength{\aboverulesep}{0pt}
    \setlength{\belowrulesep}{0pt}
    \begin{tabular}{l | >{\centering\arraybackslash}p{10cm}}
    \toprule
    \textbf{Dimension} & \textbf{Objective and Description} \\
    \midrule
    \makecell[tl]{Problem \\ Background} &
    Summarizes the task context, overall objective, and necessary background to clarify the purpose of the problem. \\
    \hline
    \makecell[tl]{Functional \\ Requirements} &
    Defines the required program behavior, including the problem objective and key operations to be performed. \\
    \hline
    \makecell[tl]{Input \\ Requirements} &
    Details the expected input data types, structure, format, ranges, and validation constraints. \\
    \hline
    \makecell[tl]{Output \\ Requirements} &
    Defines the expected output data type, format, and conditions that the generated output must satisfy. \\
    \hline
    \makecell[tl]{Test \\ Cases} &
    Extracts available examples or test cases to illustrate expected behavior and boundary conditions. \\
    \hline
    \makecell[tl]{External \\ Dependencies} &
    Identifies required external libraries, frameworks, or APIs and their usage constraints. (Optional) \\
    \hline
    \makecell[tl]{Additional \\ Notes} &
    Captures edge cases, implicit assumptions, exceptional conditions, and other constraints not covered above. (Optional) \\
    \bottomrule
    \end{tabular}
    \caption{Dimensions of Structured Specification Analyses.}    \label{tab:spec_dimensions}
\end{table}

\subsubsection{Specification-Guided Code Discrimination Dataset \texorpdfstring{$\mathcal{D}_{\text{disc}}$}{D\_disc}}
\label{sec:code_discrimination_dataset}

The dataset $\mathcal{D}_{\text{disc}}$ provides training instances for specification-guided code discrimination.
For each verified triplet $\langle p, s, c \rangle$ in $\mathcal{D}_{\text{gen}}$, we use the raw requirement $p$ and structured specification analysis $s$ as the shared context and prompt \textit{Qwen2.5-Coder-32B-Instruct} to sample ten candidate implementations conditioned on them.
Each candidate is executed on $\mathcal{T}_p$ to obtain its test pass rate, and candidate implementations with syntax errors or compilation failures are discarded.
From the remaining candidates, we construct pairwise samples with different pass rates.
Each pair is denoted as $(c^+, c^-)$, where $f(c^+,\mathcal{T}_p) > f(c^-,\mathcal{T}_p)$.
During training, the two candidate implementations are randomly ordered as $(c_A, c_B)$ to reduce position bias, forming an instance $(p, s, c_A, c_B, y)$, where $y$ labels the candidate implementation with the higher test pass rate after random ordering.

To support curriculum learning, each discrimination sample is assigned a difficulty label according to the pass-rate gap $\Delta$ between the higher-performing and lower-performing candidates.
(1) \textbf{Easy} samples correspond to $\Delta \in (0.6,1.0]$ where the programs differ substantially in their test pass rates. 
(2) \textbf{Medium} samples feature $\Delta \in (0.3,0.6]$ representing a moderate gap that often indicates partial functional differences. 
(3) \textbf{Hard} samples involve $\Delta \in (0.0,0.3]$ with a small pass-rate gap requiring much more fine-grained comparison. 
These difficulty labels subsequently guide the curriculum reinforcement learning strategy.

\subsection{Two-Stage Training Paradigm}
After data construction, SpecCoder proceeds with two-stage training, corresponding to Stage 2 and Stage 3 in Figure~\ref{figs:overview}.
The first stage warm-starts the LLM through specification-guided SFT, and the second stage further improves both generation and discrimination through curriculum dual-task GRPO.
The following subsections detail these two stages.

\subsubsection{Specification-Guided SFT}
\label{sec:sft}

As the first training stage, specification-guided SFT initializes the LLM's ability to derive structured specification analyses from raw requirements and use them to guide code generation.
For each training tuple $\langle p_i,s_i,c_i\rangle \in \mathcal{D}_{\text{gen}}$, we construct an autoregressive training sequence $X_i$ with an input part and a target part.
The input part contains the raw requirement $p_i$, enclosed by \texttt{<REQ>} and \texttt{</REQ>}.
The target part contains the structured specification analysis $s_i$, enclosed by \texttt{<ANALYZE>} and \texttt{</ANALYZE>}, followed by the implementation $c_i$, enclosed by \texttt{<CODE>} and \texttt{</CODE>}.
Under this format, the LLM learns to first produce a structured specification analysis and then generate code conditioned on both the raw requirement and the generated analysis.
We optimize the LLM by minimizing the following loss:
\begin{equation}
\mathcal{L}_{\text{SFT}} = - \frac{1}{|\mathcal{D}_{\text{gen}}|} \sum_{\langle p_i,s_i,c_i\rangle \in \mathcal{D}_{\text{gen}}} \frac{1}{\sum_{j=1}^{|X_i|} m_{i,j}} \sum_{j=1}^{|X_i|} m_{i,j} \log \pi_\theta(X_{i,j} \mid X_{i,<j}),
\end{equation}
Here, $X_{i,j}$ denotes the $j$-th token of $X_i$, and $X_{i,<j}$ denotes its preceding context.
The mask $m_{i,j} \in \{0,1\}$ determines whether each token contributes to the loss.
We set $m_{i,j}=0$ for input tokens corresponding to the raw requirement and its delimiters, and set $m_{i,j}=1$ for target tokens corresponding to the structured specification analysis, the implementation, and their target-side delimiters.
Thus, the loss is applied only to the model-generated part of the sequence.
The resulting SFT policy provides a stable initialization for the subsequent curriculum dual-task GRPO stage, where the LLM is further optimized with task-specific reward signals.

\subsubsection{Curriculum Dual-Task GRPO}
\label{sec:grpo}
After specification-guided SFT, the LLM learns to follow the analysis-before-code generation format.
\textcolor{black}{However, SFT alone provides limited optimization for strengthening the correspondence between structured specifications and code behavior across generation and implementation discrimination. Therefore, SpecCoder applies curriculum dual-task GRPO to jointly optimize specification-guided generation and discrimination using task-specific outcome rewards.}

\paragraph{\textbf{Dual-Task GRPO Reward and Advantage Design}}
SpecCoder jointly optimizes specification-guided generation and discrimination in the GRPO stage.
Since the two tasks have different output formats and reward structures, we define task-specific rewards and normalize advantages separately within each task type.
For a training instance of task type $q \in \{\mathrm{gen}, \mathrm{disc}\}$, let $\{o_{i,k}^{(q)}\}_{k=1}^{G}$ denote the $G$ sampled outputs from the current policy, and let $R_q(o_{i,k}^{(q)})$ denote the corresponding task-specific reward.
The normalized advantage is computed as:
\begin{equation}
\hat{A}_{i,k}^{(q)} =
\frac{
R_q(o_{i,k}^{(q)}) -
\operatorname{mean}\left(R_q(o_{i,1}^{(q)}), \ldots, R_q(o_{i,G}^{(q)})\right)
}{
\max\left(
\operatorname{std}\left(R_q(o_{i,1}^{(q)}), \ldots, R_q(o_{i,G}^{(q)})\right),
\sigma_{\min}^{(q)}
\right)
}.
\end{equation}
Here, $\sigma_{\min}^{(q)}$ is a task-specific variance floor, which is set to $10^{-3}$ for both generation and discrimination.
This task-wise normalization compares sampled outputs only within the same task type and training instance, preventing reward-scale differences between generation and discrimination from dominating policy updates.
We define the task-specific rewards $R_{\mathrm{gen}}$ and $R_{\mathrm{disc}}$ as follows.

For the \textbf{generation task}, let $o_g$ denote a sampled output and let $c_g$ be the implementation extracted from its code section.
The generation reward combines format compliance, compilation success, and execution-based correctness:
\begin{equation}
R_{\text{gen}}(o_g) = \alpha r_{\text{format}}^{\text{gen}} + \gamma r_{\text{compile}}^{\text{gen}} + (1-\alpha-\gamma) r_{\text{correct}}^{\text{gen}},
\end{equation}
where $\alpha=0.2$ and $\gamma=0.2$.
The format reward $r_{\text{format}}^{\text{gen}}$ is $1.0$ when both the structured specification analysis section and the code section are present, $0.75$ when only the code section is present, $0.25$ when only the structured specification analysis section is present, and $0$ otherwise.
The compilation reward $r_{\text{compile}}^{\text{gen}}$ is $1$ if the extracted implementation compiles successfully and $0$ otherwise.
The correctness reward is based on the test pass rate:
\begin{equation}
r_{\text{correct}}^{\text{gen}} =
\begin{cases}
f(c_g,\mathcal{T}_p), & \text{if } d_g=\text{Easy}, \\
\sqrt{f(c_g,\mathcal{T}_p)}, & \text{if } d_g \in \{\text{Medium}, \text{Hard}\},
\end{cases}
\end{equation}
where $d_g$ denotes the difficulty of the generation instance.

For the \textbf{discrimination task}, let $o_d$ denote a sampled output and let $\hat{y}$ be the final decision extracted from $o_d$. 
Given a discrimination instance $(p,s,c_A,c_B,y) \in \mathcal{D}_{\text{disc}}$, the discrimination reward combines a format reward and a decision correctness reward:
\begin{equation}
R_{\text{disc}}(o_d) = \lambda r_{\text{format}}^{\text{disc}} + (1-\lambda) r_{\text{decision}}^{\text{disc}},
\end{equation}
We set $\lambda=0.3$. 
The format reward $r_{\text{format}}^{\text{disc}}$ is $1.0$ when both the comparison analysis and final decision are present, $0.5$ when only one of them is present, and $0$ otherwise. 
The decision correctness reward is defined as:
\begin{equation}
r_{\text{decision}}^{\text{disc}} = \mathbb{I}[\hat{y}=y],
\end{equation}
where $y \in \{A,B\}$ denotes the candidate implementation with the higher test pass rate after random ordering.
Detailed training hyperparameters are reported in~\ref{appendix_hyper}.

\paragraph{\textbf{Progressive Curriculum Learning Strategy}}
In the curriculum dual-task GRPO stage, directly mixing all difficulty levels from the beginning may lead to unstable optimization, because easy samples provide dense but limited learning signals, whereas harder samples require more complex requirement analysis, specification-guided code generation, and fine-grained implementation comparison.
SpecCoder therefore adopts a progressive curriculum that organizes training samples according to task type and difficulty level.

For the generation task, each instance is assigned a difficulty label $d_g \in \{\text{Easy}, \text{Medium}, \text{Hard}\}$ according to the benchmark-provided difficulty annotations.
For the discrimination task, difficulty is determined by the pass rate gap $\Delta$ between the higher-performing and lower-performing candidate implementations, as defined in Section~\ref{sec:code_discrimination_dataset}.
To avoid unequal sample exposure, we keep a fixed sample-level update budget across the GRPO stage, so that the curriculum changes the training order and phase composition rather than the total exposure of each sample.
The curriculum consists of three sequential phases:
\begin{itemize}
\item \textbf{Phase 1 (Initial Alignment)}: uses Easy samples from both generation and discrimination tasks to establish basic associations between structured specification analyses and code behavior.
\item \textbf{Phase 2 (Difficulty Adaptation)}: focuses on Medium and Hard samples from both tasks to improve complex requirement analysis, specification-guided code generation, and fine-grained implementation comparison.
\item \textbf{Phase 3 (Full Consolidation)}: mixes all task types and difficulty levels to consolidate learned behaviors across the full training distribution and mitigate forgetting of easier cases.
\end{itemize}

\textcolor{black}{
Detailed training data configurations and optimization procedures are provided in Appendix~\ref{appendix_hyper}, while the complete training process is summarized in Algorithm~\ref{alg:training}.
}

\section{Experimental Design}
\label{sec:experiments_design}

\subsection{Datasets}
\label{sec:datasets}
We evaluate SpecCoder on three widely adopted competitive code generation benchmarks: APPS~\cite{hendrycks2021measuring}, CodeContests~\cite{li2022competition}, and xCodeEval~\cite{khan2024xcodeeval}. 
Following the pipeline in Section~\ref{sec:data_construction}, we construct the specification-guided generation and discrimination datasets for training. 
Table~\ref{tab:data_info} summarizes the final training and testing data distributions, while further dataset details and difficulty classifications are provided in~\ref{appendix_data_info}.

\begin{table}[htbp]
    \centering
    \small
    \renewcommand{\arraystretch}{1.0} 
    \setlength{\abovecaptionskip}{0.1cm} 
    \resizebox{0.95\linewidth}{!}{
    \begin{tabular}{l|l|c||l|l|c}
    \hline
    \multicolumn{3}{c||}{\textbf{Training Phase}} & \multicolumn{3}{c}{\textbf{Evaluation Phase}} \\
    \hline
    \textbf{Dataset} & \textbf{Phase} & \textbf{Size} & \textbf{Dataset} & \textbf{Phase} & \textbf{Size} \\
    \hline
    \multirow{2}{*}{APPS}          & SFT ($T_1:T_2=1:0$)  & 6,039 & APPS         & Testing & 500 \\
    \cline{2-3} \cline{4-6}
                                   & GRPO ($T_1:T_2=2:1$) & 5,110 & CodeContests & Testing & 165 \\
    \hline     
    \multirow{2}{*}{CodeContests} & SFT ($T_1:T_2=1:0$)  & 7,493 & xCodeEval    & Testing & 300 \\
    \cline{2-3} \cline{4-6}
                                   & GRPO ($T_1:T_2=2:1$) & 5,110 &   -          &  -      &  -  \\
    \hline
    \end{tabular}%
    }
    \caption{Statistics of the datasets. \textit{Note:} $T_1$ and $T_2$ denote the Specification-Guided Code Generation task and the Specification-Guided Code Discrimination task, respectively.}
    \label{tab:data_info}
\end{table}

\subsection{Evaluation Metrics}
Following studies~\cite{jiang2024selfplanning,tian2025ufiX}, we employ Pass@1 and AvgPassRatio to evaluate the performance of SpecCoder, capturing both exact and partial correctness.
\textcolor{black}{To quantify uncertainty, we report 95\% confidence intervals estimated via bootstrap resampling with 10,000 iterations over the test set~\cite{diciccio1996bootstrap}.}

\textbf{Pass@1.}
We focus on Pass@1 as it mirrors real-world scenarios where developers typically rely on the first generated solution~\cite{chen2024teaching,dong2024selfcollaboration,li2025codeprm}. 
For an evaluation set of $N$ tasks, the model generates a single code instance per problem with a decoding temperature of $0.2$. 
Let the indicator function $\mathbb{I}_i = 1$ if the $i$-th task passes all test cases, and $\mathbb{I}_i = 0$ otherwise. 
The metric is defined as:
\begin{equation}
Pass@1 = \frac{1}{N} \sum_{i=1}^{N} \mathbb{I}_i
\end{equation}

\textbf{AvgPassRatio.}
To provide a complementary view of partial correctness, AvgPassRatio measures the average fraction of passed test cases across the same $N$ tasks. 
For the $i$-th task, let $T_i$ denote the total number of evaluation test cases, and $t_i$ denote the number of test cases passed by the generated instance:
\begin{equation}
AvgPassRatio = \frac{1}{N} \sum_{i=1}^{N} \frac{t_i}{T_i}
\end{equation}

\subsection{Baselines}

To compare SpecCoder with representative code generation methods, we consider two broad categories of baselines: training-free and training-based methods. 
The training-free baselines include prompt-based approaches, such as SCoT~\cite{li2025SCoT}, Self-Planning~\cite{jiang2024selfplanning}, $\mu$FiX~\cite{tian2025ufiX} and \textcolor{black}{ArchCode}~\cite{han2024archcode}.
The training-based baselines include CodeRL+~\cite{jiang2025coderl+}, CodeDPO~\cite{zhang2025codedpo}, CodePRM~\cite{li2025codeprm}, and \textcolor{black}{FixAudit}~\cite{tang2026fixaudit} which optimize code generation through reinforcement learning, preference optimization, or reward modeling.
To further evaluate the generalization and compatibility of SpecCoder as a backbone model, we extend our study to four representative training-free agent-based frameworks: AgentCoder~\cite{huang2023agentcoder}, MetaGPT~\cite{hong2023metagpt}, PairCoder~\cite{zhang2024paircoder}, and Specine~\cite{tian2025aligning}. 
This setting allows us to examine whether the specification-aware generation capability learned by SpecCoder can benefit downstream agent workflows across different agent architectures.
More
details on these approaches can be found in their papers.

\subsection{Implementation Details}

All training and evaluation tasks are executed on two NVIDIA A100 SXM4 GPUs with 80GB memory. SpecCoder follows a two-stage training scheme. In the first stage, we perform SFT on \textit{Qwen2.5-7B-Coder-Instruct}~\cite{hui2024qwen2} with LoRA fine tuning~\cite{hu2021lora} using LLaMA Factory~\cite{zheng2024llamafactory}. 
In the second stage, we perform curriculum dual-task GRPO~\cite{shao2024deepseekmath} using SkyRL~\cite{liu2025skyrlsql}.
For fair comparison, all trainable methods use the same backbone model, training split, evaluation protocol, hardware, and comparable optimization budget. 
Training-free baselines, including prompt-based and agent-based methods, use the same backbone model and decoding settings without parameter updates. 
Training-based baselines are trained on the same programming problems and test suites, with training signals following their original designs. 
Detailed data distributions, hardware statistics, and hyperparameter configurations are provided in~\ref{appendix_hyper}.

\begin{table*}[htbp]
  \centering
  \setlength{\extrarowheight}{3pt} 
  \resizebox{\textwidth}{!}{%
  \begin{tabular}{l|c|c|c|c|c|c} 
    \hline
    \multirow{3}{*}{\textbf{Technique}} & \multicolumn{4}{c|}{\textbf{ID}} & \multicolumn{2}{c}{\textbf{OOD}} \\
    \cline{2-7} 
              & \multicolumn{2}{c|}{\textbf{APPS}} & \multicolumn{2}{c|}{\textbf{CodeContests}} & \multicolumn{2}{c}{\textbf{xCodeEval}} \\
    \cline{2-7}
              & \textbf{P@1} & \textbf{Avg.P} & \textbf{P@1} & \textbf{Avg.P} & \textbf{P@1} & \textbf{Avg.P} \\
    \hline 
    \multicolumn{7}{l}{\textit{\textbf{Training-free Prompting Baselines}}} \\
    \hline
    Zero-shot       & \mbox{$0.166 \pm 0.033$} & \mbox{$0.271 \pm 0.033$} & \mbox{$0.030 \pm 0.027$} & \mbox{$0.145 \pm 0.037$} & \mbox{$0.100 \pm 0.032$} & \mbox{$0.204 \pm 0.038$} \\
    SCoT            & \mbox{$0.168 \pm 0.033$} & \mbox{$0.298 \pm 0.033$} & \mbox{$0.036 \pm 0.027$} & \mbox{$0.114 \pm 0.037$} & \mbox{$0.097 \pm 0.032$} & \mbox{$0.250 \pm 0.038$} \\
    Self-Planning   & \mbox{$0.168 \pm 0.030$} & \mbox{$0.278 \pm 0.031$} & \mbox{$0.049 \pm 0.033$} & \mbox{$0.109 \pm 0.040$} & \mbox{$0.083 \pm 0.030$} & \mbox{$0.233 \pm 0.037$} \\
    $\mu$Fix        & \mbox{$0.170 \pm 0.031$} & \mbox{$0.299 \pm 0.032$} & \mbox{$0.055 \pm 0.033$} & \mbox{$0.157 \pm 0.037$} & \mbox{$0.133 \pm 0.025$} & \mbox{$0.279 \pm 0.030$} \\
    \textcolor{black}{ArchCode}  & \mbox{\textcolor{black}{$0.174 \pm 0.031$}} & \mbox{\textcolor{black}{$0.274 \pm 0.032$}} & \mbox{\textcolor{black}{$0.067 \pm 0.025$}} & \mbox{\textcolor{black}{$0.145 \pm 0.041$}} & \mbox{\textcolor{black}{$0.103 \pm 0.032$}} & \mbox{\textcolor{black}{$0.189 \pm 0.036$}}  \\
    \hline 
    \multicolumn{7}{l}{\textit{\textbf{Training-based Fine-tuning Baselines}}} \\
    \hline
    CodeRL+         & \mbox{$0.182 \pm 0.039$} & \mbox{$0.339 \pm 0.036$} & \mbox{$0.049 \pm 0.033$} & \mbox{$0.105 \pm 0.039$} & \mbox{$0.093 \pm 0.028$} & \mbox{$0.175 \pm 0.034$} \\
    CodeDPO         & \mbox{$0.178 \pm 0.035$} & \mbox{$0.332 \pm 0.034$} & \mbox{$0.061 \pm 0.033$} & \mbox{$0.144 \pm 0.044$} & \mbox{$0.073 \pm 0.031$} & \mbox{$0.171 \pm 0.035$} \\
    CodePRM         & \mbox{$\mathbf{0.186} \pm 0.033$} & \mbox{$0.340 \pm 0.033$} & \mbox{$0.073 \pm 0.037$} & \mbox{$0.166 \pm 0.035$} & \mbox{$0.103 \pm 0.033$} & \mbox{$0.207 \pm 0.039$} \\
    \textcolor{black}{FixAudit}  & \mbox{\textcolor{black}{$0.184 \pm 0.035$}} & \mbox{\textcolor{black}{$0.254 \pm 0.032$}} & \mbox{\textcolor{black}{$\mathbf{0.091} \pm 0.042$}} & \mbox{\textcolor{black}{$0.162 \pm 0.046$}} & \mbox{\textcolor{black}{$0.143 \pm 0.040$}} & \mbox{\textcolor{black}{$0.279 \pm 0.044$}} \\
    \hline
    \multicolumn{7}{l}{\textit{\textbf{Ours}}} \\
    \hline
    \textbf{SpecCoder} & \mbox{$\mathbf{0.186} \pm \mathbf{0.028}$} & \mbox{$\mathbf{0.346} \pm \mathbf{0.030}$} & \mbox{$\mathbf{0.091} \pm \mathbf{0.023}$} & \mbox{$\mathbf{0.207} \pm \mathbf{0.033}$} & \mbox{$\mathbf{0.147} \pm \mathbf{0.026}$} & \mbox{$\mathbf{0.324} \pm \mathbf{0.033}$} \\
    \hline  
  \end{tabular}
  }
  \caption{\textcolor{black}{Performance comparison of various code generation techniques utilizing the base \textit{Qwen2.5-7B-Coder-Instruct}, measured by Pass@1 and AvgPassRatio. Results are reported with 95\% confidence intervals, where P@1 and Avg.P denote Pass@1 and AvgPassRatio, respectively.}}
  \label{tab:main_results} 
\end{table*}

\section{Results and Analysis}

\subsection{Effectiveness of the SpecCoder Framework}

\textbf{Setup.} To establish a unified experimental foundation, we use the open-source \textit{Qwen2.5-7B-Coder-Instruct} as the baseline backbone.
\textcolor{black}{In this subsection, we evaluate standalone code generation by assessing ID performance on APPS and CodeContests along with OOD generalization on xCodeEval. Across these evaluations, we compare SpecCoder against a standard Zero-shot approach and the previously introduced training-free and training-based baselines.}
All methods generate a single solution with a decoding temperature of $0.2$. 
\textcolor{black}{All metric scores are reported with 95\% bootstrap confidence intervals estimated over test instances.}

\textbf{Results and Analysis.}
Table~\ref{tab:main_results} summarizes the performance of standalone code generation methods across the three datasets.
For ID evaluation, SpecCoder achieves strong performance on both APPS and CodeContests. On APPS, SpecCoder obtains a Pass@1 of $0.186$ and an AvgPassRatio of $0.346$. Compared with the strongest baseline CodePRM which achieves $0.186$ and $0.340$, SpecCoder matches its Pass@1 while further improving AvgPassRatio. 
\textcolor{black}{On CodeContests, SpecCoder and FixAudit achieve the same Pass@1 of $0.091$, while SpecCoder achieves a higher AvgPassRatio of $0.207$ compared with $0.162$ for FixAudit.
For OOD evaluation on xCodeEval, SpecCoder achieves a Pass@1 of $0.147$, compared with $0.143$ for FixAudit and $0.133$ for $\mu$Fix.
Furthermore, SpecCoder achieves the highest AvgPassRatio of $0.324$, delivering an improvement of $4.5$ percentage points over the best baseline.}

\begin{tcolorbox}[
colback=gray!5!white,
colframe=gray!95!black,
fonttitle=\bfseries,
arc=2.8pt,
boxrule=0.9pt,
enhanced,
boxsep=1.5pt,
left=1.5pt, right=1.5pt,
top=1.5pt, bottom=1.5pt,
before upper={\setstretch{0.9}}
]
\textbf{Conclusion 1}: \textcolor{black}{SpecCoder achieves competitive or superior standalone code generation performance across the evaluated ID and OOD benchmarks, supporting the effectiveness of the proposed specification-aware two-stage training framework.}
\end{tcolorbox}

\begin{figure}[htbp]
    \centering
    \includegraphics[width=0.9\linewidth]{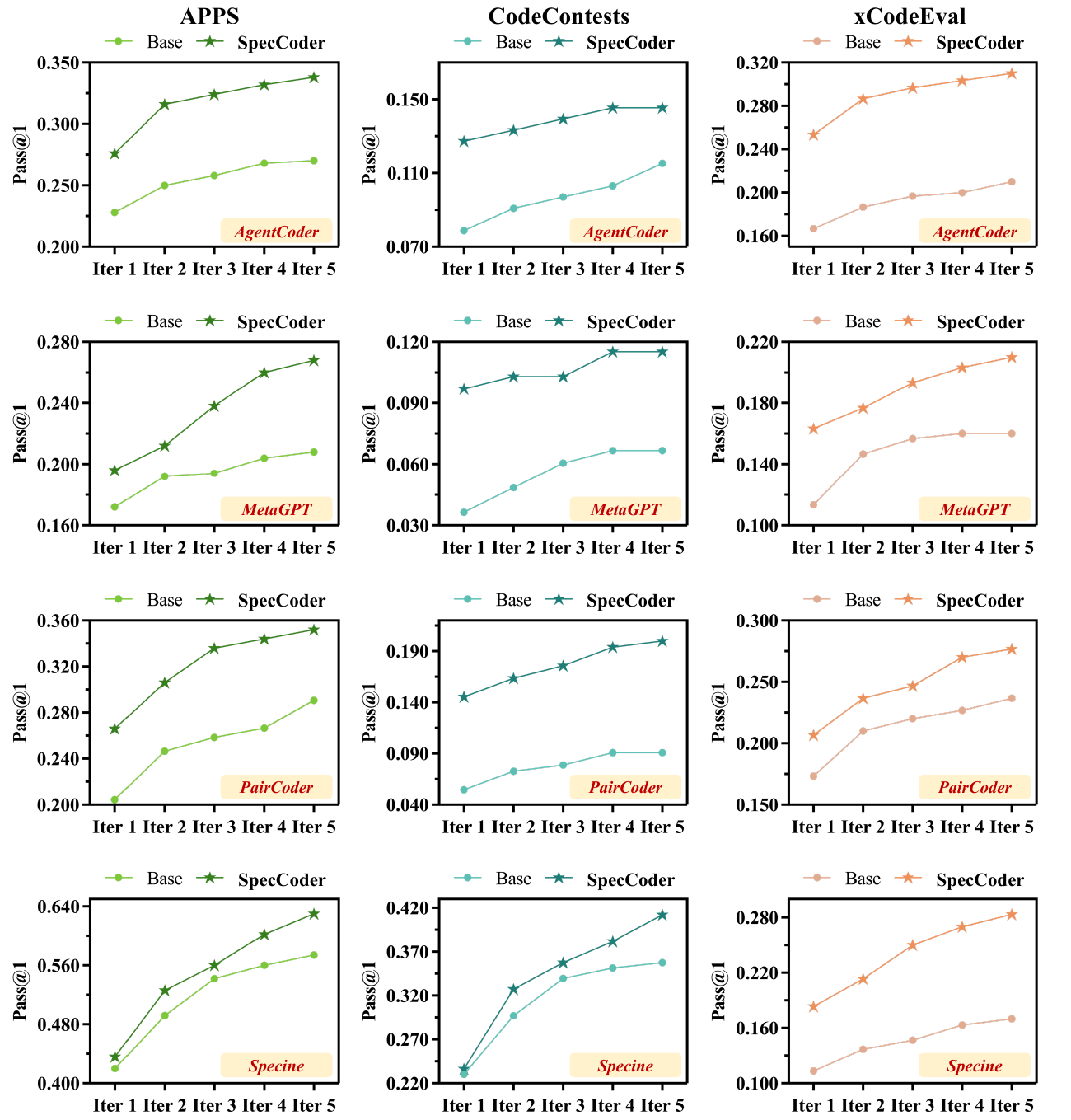}
    % \vspace{-0.5cm}
    \caption{Performance trends in terms of Pass@1 of agent-based code generation methods over five iterations across three datasets. Each row corresponds to a specific agent framework, comparing the base \textit{Qwen2.5-7B-Coder-Instruct} with its SpecCoder-enhanced variant.}
    \label{figs:agent_iteration}
\end{figure}

\subsection{Generalization as an Agent Backbone}

\textbf{Setup.}
To evaluate whether SpecCoder can generalize beyond standalone code generation, we use it as the backbone model of representative training-free agent-based workflows, including AgentCoder, MetaGPT, PairCoder, and Specine.
Specifically, we replace the original \textit{Qwen2.5-7B-Coder-Instruct} backbone in each workflow with the LLM trained by SpecCoder.
We evaluate agent performance in both ID and OOD settings over five interactive rounds.

\begin{table*}[htbp]
  \centering
  \setlength{\extrarowheight}{3pt} 
  \setlength{\tabcolsep}{3pt} 
  \resizebox{\textwidth}{!}{%
  \begin{tabular}{c|l|c|c|c|c|c|c}
    \hline
    \multicolumn{2}{c|}{\multirow{3}{*}{\textbf{Technique}}} & \multicolumn{4}{c|}{\textbf{ID}} & \multicolumn{2}{c}{\textbf{OOD}} \\
    \cline{3-8}
    \multicolumn{2}{c|}{} & \multicolumn{2}{c|}{\textbf{APPS}} & \multicolumn{2}{c|}{\textbf{CodeContests}} & \multicolumn{2}{c}{\textbf{xCodeEval}} \\
    \cline{3-8}
    \multicolumn{2}{c|}{} & \textbf{P@1} & \textbf{Avg.P} & \textbf{P@1} & \textbf{Avg.P} & \textbf{P@1} & \textbf{Avg.P} \\
    \hline
    \multirow{2}{*}{AgentCoder} & Base  & $0.270 \pm 0.039$ & $0.501 \pm 0.033$ & $0.115 \pm 0.048$ & $0.236 \pm 0.052$ & $0.210 \pm 0.045$ & $0.431 \pm 0.043$ \\
    \cline{2-8}
    ~ & \textbf{SpecCoder} & $0.338 \pm 0.041$ & $0.541 \pm 0.034$ & $0.146 \pm 0.052$ & $0.328 \pm 0.056$ & $0.310 \pm 0.052$ & $0.563 \pm 0.041$ \\
    \hline
    \multirow{2}{*}{MetaGPT} & Base  & $0.208 \pm 0.036$ & $0.393 \pm 0.033$ & $0.067 \pm 0.039$ & $0.204 \pm 0.048$ & $0.160 \pm 0.042$ & $0.375 \pm 0.041$ \\
    \cline{2-8}
    ~ & \textbf{SpecCoder} & $0.268 \pm 0.039$ & $0.486 \pm 0.033$ & $0.115 \pm 0.048$ & $0.293 \pm 0.052$ & $0.210 \pm 0.047$ & $0.469 \pm 0.041$ \\
    \hline
    \multirow{2}{*}{PairCoder} & Base  & $0.291 \pm 0.041$ & $0.481 \pm 0.036$ & $0.091 \pm 0.045$ & $0.254 \pm 0.052$ & $0.237 \pm 0.048$ & $0.423 \pm 0.046$ \\
    \cline{2-8}
    ~ & \textbf{SpecCoder} & $0.352 \pm 0.042$ & $0.574 \pm 0.033$ & $0.200 \pm 0.064$ & $0.377 \pm 0.061$ & $0.277 \pm 0.050$ & $0.542 \pm 0.041$ \\
    \hline
    \multirow{2}{*}{Specine} & Base  & $0.596 \pm 0.042$ & $0.700 \pm 0.036$ & $0.376 \pm 0.073$ & $0.426 \pm 0.071$ & $0.170 \pm 0.043$ & $0.243 \pm 0.043$ \\
    \cline{2-8}
    ~ & \textbf{SpecCoder} & $0.656 \pm 0.041$ & $0.749 \pm 0.033$ & $0.424 \pm 0.073$ & $0.456 \pm 0.073$ & $0.283 \pm 0.050$ & $0.368 \pm 0.048$ \\
    \hline
  \end{tabular}%
  }
  \caption{\textcolor{black}{Performance comparison of different agent-based techniques using \textit{Qwen2.5-7B-Coder-Instruct} at iteration 5, measured by Pass@1 and AvgPassRatio. Results are reported with 95\% confidence intervals, where P@1 and Avg.P denote Pass@1 and AvgPassRatio, respectively.}}
  \label{tab:agent_results}
\end{table*}

\textbf{Results and Analysis.} 
% Table~\ref{tab:agent_results} reports the final Pass@1 and AvgPassRatio of four agent-based workflows after five rounds, while Figure~\ref{figs:agent_iteration} shows the Pass@1 trajectories across iterations.
As Table~\ref{tab:agent_results} shows, SpecCoder consistently improves agent-based workflows on both ID and OOD benchmarks. On APPS, PairCoder achieves the largest gain, with Pass@1 increasing by 6.1 percentage points over the original backbone. On the more challenging CodeContests benchmark, replacing the backbone with SpecCoder improves Pass@1 and AvgPassRatio by up to 10.9 and 12.3 percentage points, respectively. On xCodeEval, SpecCoder also shows strong OOD generalization, with Specine reaching 28.3\% Pass@1 and 36.8\% AvgPassRatio, corresponding to gains of 11.3 and 12.5 percentage points. 
\textcolor{black}{The results after five iterations consistently show higher performance for SpecCoder-based agents than for their corresponding base-model counterparts across the evaluated frameworks. These results demonstrate that SpecCoder serves as an effective backbone for diverse agent architectures.}

Figure~\ref{figs:agent_iteration} further shows that SpecCoder-based agents consistently achieve higher Pass@1 than agents using the original LLM at each refinement round, while also showing gradual performance improvements across iterations. 
\textcolor{black}{This observation confirms that the performance lead of SpecCoder is consistently maintained throughout the entire multi round interactive process.}

\begin{tcolorbox}[
colback=gray!5!white,
colframe=gray!95!black,
fonttitle=\bfseries,
arc=2.8pt,
boxrule=0.9pt,
enhanced,
boxsep=1.5pt,
left=1.5pt, right=1.5pt,
top=1.5pt, bottom=1.5pt,
before upper={\setstretch{0.9}}
]
\textbf{Conclusion 2}: SpecCoder provides a stronger backbone for training-free agent-based code generation workflows across both ID and OOD benchmarks. 
\textcolor{black}{he experimental results confirm that replacing the base model with SpecCoder directly elevates the final performance of various multi-step agent workflows without requiring any modifications to their external workflow design.}
\end{tcolorbox}

\begin{figure}[!t]
    \centering
    \includegraphics[width=\linewidth]{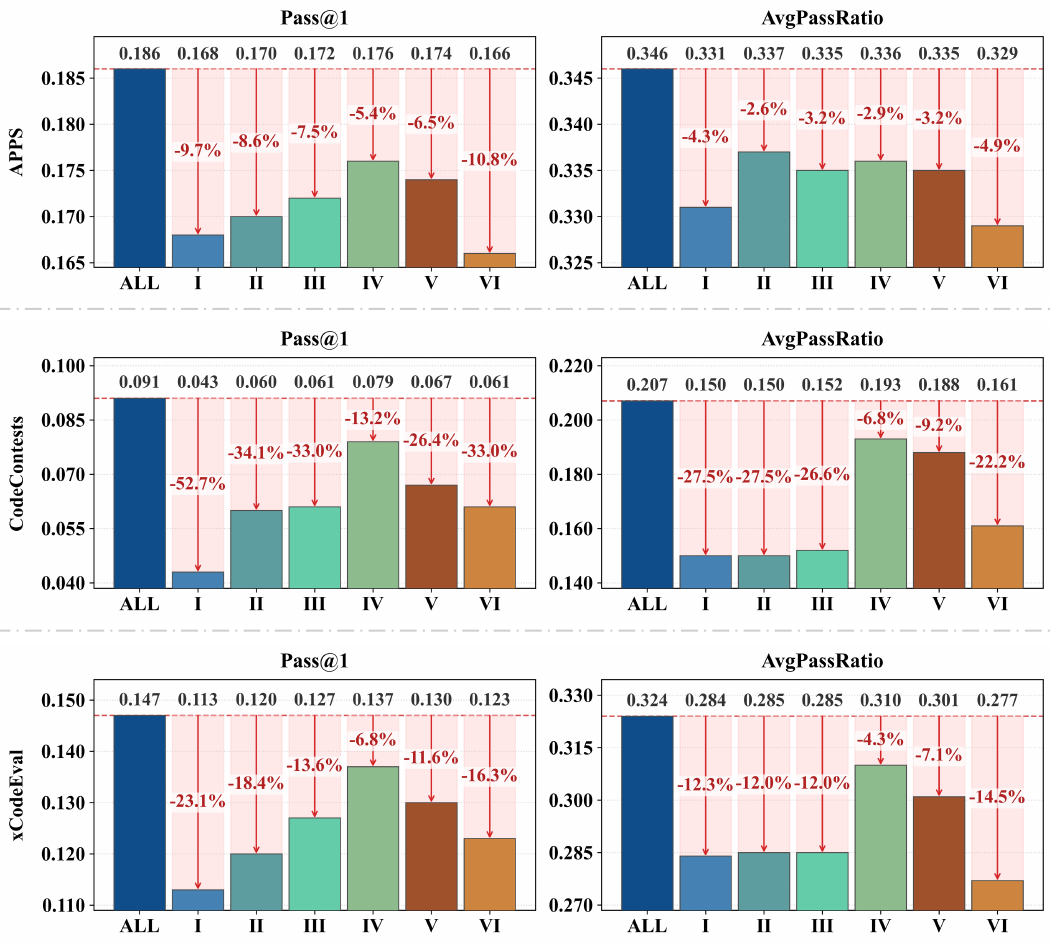}
    \caption{Ablation study results of SpecCoder in terms of Pass@1 and AvgPassRatio across three benchmarks. The red shaded areas and arrows quantify the performance drop of each variant relative to the full model. Configuration labels: \textbf{I}~w/o SFT, \textbf{II}~w/o GRPO, \textbf{III}~w/o discrimination task, \textbf{IV}~w/o curriculum, \textbf{V}~generation-first sequential, \textbf{VI}~discrimination-first sequential.
    }
    \label{figs:ablation_results}
\end{figure}

\subsection{Ablation Study}

\textcolor{black}{
To verify the contribution of each design in SpecCoder we conduct ablation experiments across both ID and OOD benchmarks by dividing the variants into three main groups training component, optimization strategy and intermediate representation. Unless stated otherwise all variants share the same base model data splits evaluation protocols and compute budgets.
}

\subsubsection{\textcolor{black}{Training Component}}
\label{ablation:component}
\textbf{Setup.}
\textcolor{black}{This section} evaluates whether the main training components in the specification-aware two-stage training framework are necessary for effective code generation.
The specific ablation designs are as follows:
\begin{itemize}
\item \textbf{Variant I (w/o SFT)} removes the specification-guided SFT stage and applies curriculum dual-task GRPO directly to the base model.
\item \textbf{Variant II (w/o GRPO)} removes the curriculum dual-task GRPO stage and relies only on specification-guided SFT.
\item \textbf{Variant III (w/o Discrimination Task)} removes the specification-guided code discrimination task and optimizes only the generation task during GRPO.
\end{itemize}

\textbf{Results and Analysis.}
The overall ablation results are illustrated in Figure~\ref{figs:ablation_results}. 
\textcolor{black}{The training component ablation reveals that removing any of the three key components leads to noticeable performance degradation. 
Specifically, Variant I applies GRPO directly without supervised initialization, reducing Pass@1 on CodeContests from $0.091$ to $0.043$. This severe $52.7$ percent drop suggests the model relies on supervised fine tuning to learn how to format and utilize structured specification analyses. 
Conversely, Variant II removes GRPO and causes a $34.1$ percent decrease in Pass@1. This mutual degradation confirms that SFT provides the essential structural initialization, while GRPO further optimizes generation quality beyond supervised data.}
\textbf{Variant III} verifies the importance of the specification-guided code discrimination task.
When this task is removed, GRPO optimizes only the generation objective, resulting in weaker performance.
This indicates that generation alone does not explicitly train the LLM to judge whether an implementation follows the stated specification-level understanding.
By comparing candidate implementations under the same raw requirement and structured specification analysis, the discrimination task helps the model associate specification satisfaction with concrete code behavior.

\begin{tcolorbox}[
colback=gray!5!white,
colframe=gray!95!black,
fonttitle=\bfseries,
arc=2.8pt,
boxrule=0.9pt,
enhanced,
boxsep=1.5pt,
left=1.5pt, right=1.5pt,
top=1.5pt, bottom=1.5pt,
before upper={\setstretch{0.9}}
]
\textbf{Conclusion 3:} Each training component contributes to the overall effectiveness of SpecCoder, as removing any component leads to performance degradation. Specification-guided SFT initializes the capacity of the model to derive structured specification analyses and generate code conditioned on them. Dual-task GRPO \textcolor{black}{provides the essential initialization for structured generation while the discrimination task acts as a powerful auxiliary signal during dual task GRPO to further maximize overall code quality.}
\end{tcolorbox}

\subsubsection{\textcolor{black}{Optimization Strategy}}

\textbf{Setup.} \textcolor{black}{This experiment examines how the difficulty based curriculum scheduling and the task optimization order affect dual task GRPO.}
We design three variants to modify the curriculum schedule or the task optimization order:
\begin{itemize}
\item \textbf{Variant IV (w/o Curriculum)} removes the difficulty-based curriculum and trains on samples from all difficulty levels simultaneously.
\item \textbf{Variant V (Gen-First Sequential)} replaces joint dual-task optimization with sequential training, where the generation task is optimized before the discrimination task.
\item \textbf{Variant VI (Disc-First Sequential)} replaces joint dual-task optimization with sequential training, where the discrimination task is optimized before the generation task.
\end{itemize}

\textbf{Results and Analysis.}
\textcolor{black}{Figure~\ref{figs:ablation_results} illustrates the results of the optimization strategy ablation. 
For \textbf{Variant IV} removing the curriculum reduces Pass@1 on CodeContests from $0.091$ to $0.079$ which reveals a drop on more complex benchmarks.
This demonstrates that curriculum learning provides a stable optimization path by establishing reliable initial signals before introducing complex requirements.}
\textbf{Variant V} and \textbf{Variant VI} replace joint dual-task optimization with sequential training.
Both variants underperform the full model, indicating that optimizing generation and discrimination separately is less effective than jointly optimizing them.
Sequential training may overemphasize one task before the other has established a stable learning signal.
In contrast, joint optimization allows generation and discrimination to reinforce each other throughout GRPO.
The generation task improves the model's ability to construct and use structured specification analyses for code synthesis, while the discrimination task strengthens its ability to relate specification-level understanding to concrete code behavior.

\begin{tcolorbox}[
colback=gray!5!white,
colframe=gray!95!black,
fonttitle=\bfseries,
arc=2.8pt,
boxrule=0.9pt,
enhanced,
boxsep=1.5pt,
left=1.5pt, right=1.5pt,
top=1.5pt, bottom=1.5pt,
before upper={\setstretch{0.9}}
]
\textbf{Conclusion 4}: The optimization strategy ablation demonstrates that both curriculum learning and joint dual-task optimization are important \textcolor{black}{for the overall performance of SpecCoder. The difficulty based curriculum provides SpecCoder with a stable and progressive optimization path for complex coding requirements while joint training ensures the generation and discrimination tasks mutually reinforce the specification-aware learning process.}
\end{tcolorbox}

\subsubsection{\textcolor{black}{Intermediate Representation}}

\textbf{Setup.} 
\textcolor{black}{
This experiment further investigates whether the proposed structured specifications provide advantages over other intermediate representations under matched training conditions.
We select three representative baselines, and all intermediate representations are generated by \textbf{\textit{GPT-4o}} for the same training problems.
Then apply the same LLM filtering rules used by SpecCoder to construct high-quality training data.
During GRPO, these three variants optimize only the generation task and use the same generation rewards and training budget as the generation-only SpecCoder variant (w/o Disc), corresponding to Variant III in Section~\ref{ablation:component}.
The three baseline representations are defined as follows:}
\textcolor{black}{
\begin{itemize}
    \item \textbf{Code-Only} directly formats the training data to translate raw requirements into code without any intermediate representation, serving as the standard direct generation baseline~\cite{chen2021evaluating}.
    \item \textbf{Free-Form} structures the training data to include an unconstrained natural-language rationale prior to code generation, inspired by conventional Chain-of-Thought~\cite{wei2022chain}. This examines whether the gains arise from adding general reasoning before code generation.
    \item \textbf{Planning-CoT} formats the training sequences to first formulate a stepwise implementation plan and subsequently generate the code~\cite{jiang2024selfplanning}, controlling for the benefits of structured intermediate planning.
\end{itemize}
}

\begin{figure}[!t]
    \centering
    \includegraphics[width=\linewidth]{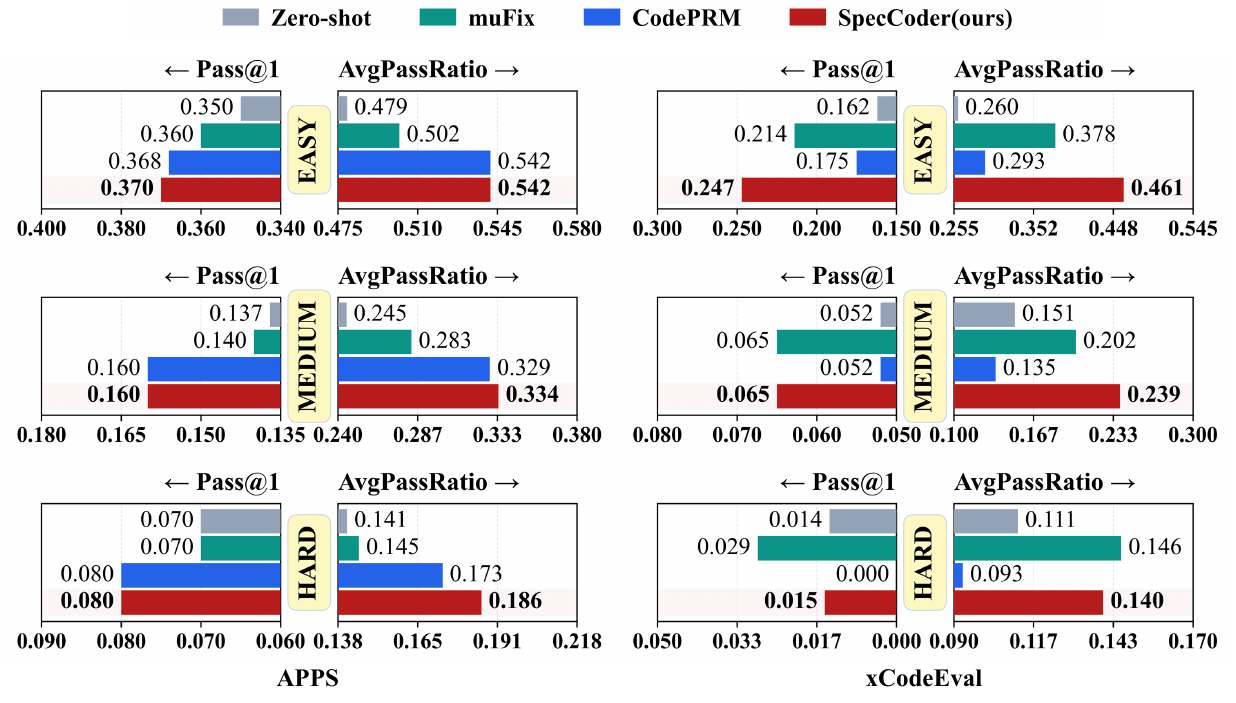}
    \caption{Performance comparison in terms of Pass@1 and AvgPassRatio on APPS and xCodeEval, across Easy, Medium, and Hard difficulty levels.}
    \label{figs:depth_analysis}
\end{figure}

\begin{figure}[!t]
    \centering
    \includegraphics[width=\linewidth]{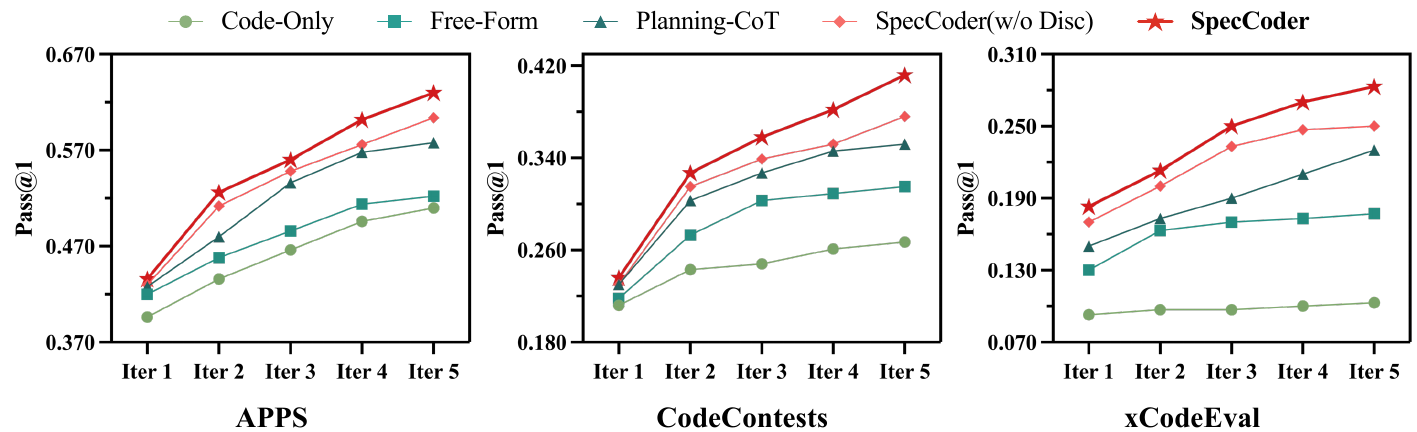}
    \caption{\textcolor{black}{Performance comparison in terms of Pass@1 for different intermediate representations across three datasets over iterations 1–5. All variants are based on the \textit{Qwen2.5-7B-Coder-Instruct} model.}}
    \label{figs:rebuttal_specine_iter}
\end{figure}

\textcolor{black}{\textbf{Results and Analysis.}}
\textcolor{black}{We report Pass@1 and average output length in tokens for standalone generation in Table~\ref{tab:representation_results} alongside Pass@1 over five Specine iterations in Figure~\ref{figs:rebuttal_specine_iter}.}

\textcolor{black}{
As shown in Table~\ref{tab:representation_results}, SpecCoder (w/o Disc) achieves the best or tied-best performance among all generation-only variants. 
On the ID benchmarks, it achieves $0.172$ on APPS and $0.061$ on CodeContests, outperforming or matching the three baselines. 
Its advantage is more evident on the OOD xCodeEval benchmark where it achieves 0.127 and outperforms Code Only Free Form and Planning CoT by $4.7$ $2.4$ and $3.4$ percentage points respectively. Building upon this generation foundation the full SpecCoder further elevates the scores to $0.186$ on APPS $0.091$ on CodeContests and $0.147$ on xCodeEval through joint generation and discrimination training
Since different intermediate representations produce outputs of different lengths, we also report their average token counts. 
SpecCoder (w/o Disc) generates only 126.5 and 155.7 more tokens than Free-Form and Planning-CoT, respectively, but improves Pass@1 on xCodeEval by 2.4 and 3.4 percentage points. The full SpecCoder adds only 14.7 tokens and further improves performance by 1.4, 3.0, and 2.0 percentage points across the three datasets. These results show that the gains do not come solely from longer outputs, but mainly from organizing requirement constraints through structured specifications and aligning requirements with code through two-stage training.
}

\textcolor{black}{
To further examine the role of different intermediate representations in multi-round code generation, we select Specine, a representative code generation agent framework, and evaluate it with all five representation variants as backbones.
As shown in Figure~\ref{figs:rebuttal_specine_iter}, the Pass@1 of all models increases across iterations. However, the later gains of Code-Only, Free-Form, and Planning-CoT gradually slow down, while the models using structured specifications maintain higher performance. At Iter 5, the full SpecCoder outperforms the best non-structured representation by 5.2, 6.0, and 5.3 percentage points on APPS, CodeContests, and xCodeEval, respectively. SpecCoder (w/o Disc) also leads by 2.6, 2.4, and 2.0 percentage points. These results show that structured specifications explicitly organize functional requirements and constraints, providing a consistent reference for iterative generation, checking, and repair. The further advantage of the full SpecCoder suggests that two-stage training strengthens the alignment between specifications and code behavior, allowing the agent to use iterative feedback more effectively.
}

\begin{table*}[!t]
  \centering
  \footnotesize 
  % \color{blue} 
  \setlength{\extrarowheight}{3pt} 
  \setlength{\tabcolsep}{8pt} 
  \begin{tabular}{l|c|c|c|c}
    \hline
    \multirow{2}{*}{\textbf{IR}} & \multicolumn{2}{c|}{\textbf{ID}} & \textbf{OOD} & \multirow{2}{*}{\textbf{\makecell{Avg. Tokens}}} \\
    \cline{2-4}
    & \textbf{APPS} & \textbf{CodeContests} & \textbf{xCodeEval} & \\
    \hline
    Code-Only & $0.166 \pm 0.029$ & $0.055 \pm 0.033$ & $0.080 \pm 0.032$ & 277.9 \\
    \hline
    Free-Form & $0.170 \pm 0.033$ & $0.060 \pm 0.039$ & $0.103 \pm 0.035$ & 568.2 \\
    \hline
    Planning-CoT & $0.168 \pm 0.031$ & $0.061 \pm 0.039$ & $0.093 \pm 0.033$ & 539.0 \\
    \hline
    SpecCoder (w/o Disc) & $0.172 \pm 0.032$ & $0.061 \pm 0.030$ & $0.127 \pm 0.028$ & 694.7 \\
    \hline
    \textbf{SpecCoder} & $\mathbf{0.186 \pm 0.028}$ & $\mathbf{0.091 \pm 0.023}$ & $\mathbf{0.147 \pm 0.026}$ & \textbf{709.4} \\
    \hline
  \end{tabular}%
  \caption{\textcolor{black}{Performance comparison of different intermediate representations in terms of Pass@1 and average output tokens. Results are reported with 95\% confidence interval}}
  \label{tab:representation_results}
\end{table*}

\begin{tcolorbox}[
colback=gray!5!white,
colframe=gray!95!black,
fonttitle=\bfseries,
arc=2.8pt,
boxrule=0.9pt,
enhanced,
boxsep=1.5pt,
left=1.5pt, right=1.5pt,
top=1.5pt, bottom=1.5pt,
before upper={\setstretch{0.9}}
]
\textbf{Conclusion 5:} \textcolor{black}{
Under matched training conditions, structured specifications achieve better overall standalone and agent-based performance than Code-Only, Free-Form, and Planning-CoT representations. These results suggest that explicitly organizing requirement information provides more effective intermediate guidance for code generation.
}
\end{tcolorbox}

\section{In-depth Analysis}

\subsection{Performance across Problem Difficulties}

\textbf{Training-based Standalone Model Performance.} 
Figure~\ref{figs:depth_analysis} evaluates standalone code generation performance across different problem difficulties on APPS and xCodeEval.
As task difficulty increases, all methods show performance degradation, while SpecCoder maintains stronger overall robustness, especially in AvgPassRatio.
Specifically, on Easy tasks, SpecCoder clearly outperforms baselines on xCodeEval, achieving a Pass@1 of $0.247$ and an AvgPassRatio of $0.461$.
On Medium tasks, SpecCoder obtains the highest AvgPassRatio of $0.171$, although its Pass@1 of $0.043$ is slightly lower than CodePRM's $0.057$.
This indicates that SpecCoder improves broader test-case coverage by aligning structured specification analyses with code behavior, even when fewer generated code implementations pass all tests.
On Hard tasks, where Pass@1 approaches zero for most methods, SpecCoder still achieves the leading AvgPassRatio of $0.186$ on APPS.
These results suggest that specification-aware training helps the LLM better preserve partial requirement satisfaction under increasing task complexity.

\begin{figure}[!t]
    \centering
    \includegraphics[width=\linewidth]{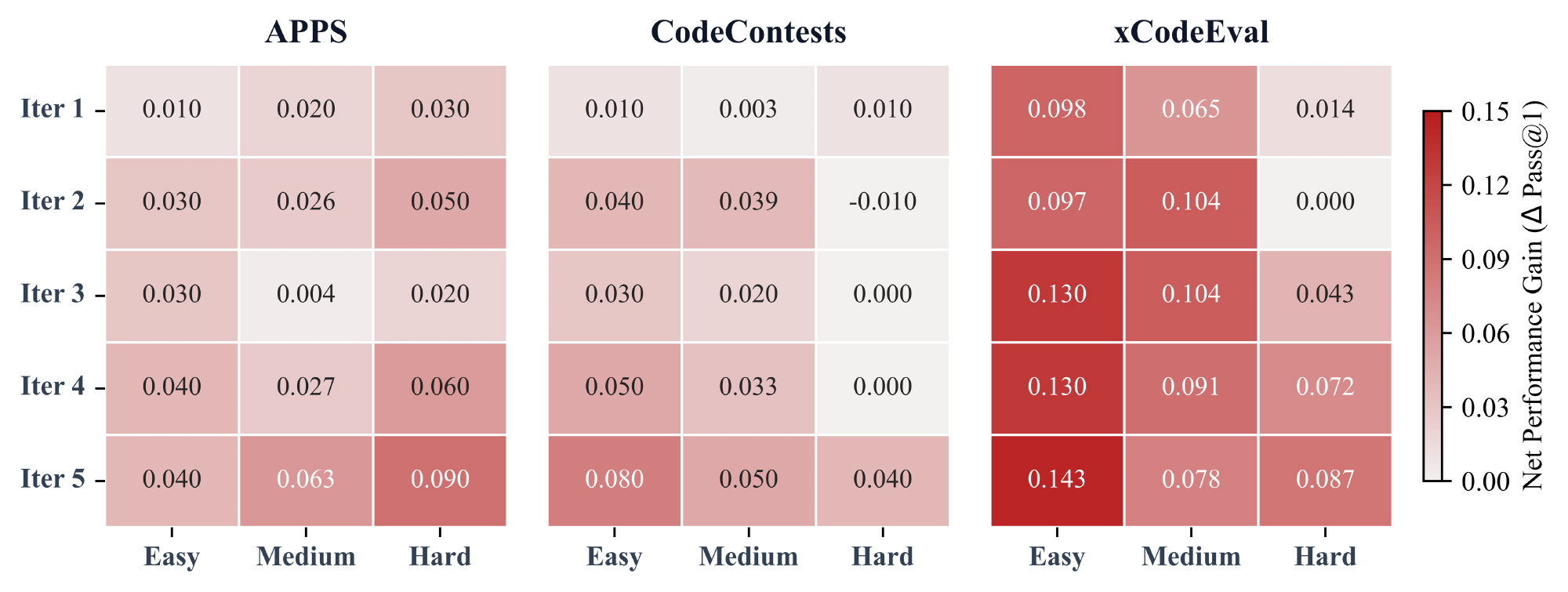}
    \caption{Heatmap of Pass@1 improvement achieved by replacing the base \textit{Qwen2.5-7B-Coder-Instruct} with SpecCoder within the Specine framework across iterations 1–5 for different difficulty levels of each dataset.}
    \label{figs:agent_analysis}
\end{figure}

\textbf{Training-free Agent-based Performance.} 
Figure~\ref{figs:agent_analysis} further evaluates SpecCoder as the backbone of a training-free agent-based self-repair framework over five interaction rounds.
Compared with the vanilla backbone, SpecCoder consistently yields larger Pass@1 gains across iterations.
For example, on CodeContests Easy tasks, the net improvement increases from $0.010$ in the first round to $0.080$ in the fifth round, while xCodeEval Easy tasks obtain a fifth-round gain of $0.143$.
This trend also holds on harder tasks.
APPS Hard tasks improve from $0.440$ to $0.530$, yielding a net gain of $0.090$, and xCodeEval Hard tasks increase from $0.174$ to $0.261$.
These results indicate that SpecCoder provides a stronger specification-aware backbone, enabling training-free agent-based workflows to start from better intermediate outputs and achieve more effective iterative repair.

\begin{tcolorbox}[
colback=gray!5!white,
colframe=gray!95!black,
fonttitle=\bfseries,
arc=2.8pt,
boxrule=0.9pt,
enhanced,
boxsep=1.5pt,
left=1.5pt, right=1.5pt,
top=1.5pt, bottom=1.5pt,
before upper={\setstretch{0.9}}
]
\textbf{Conclusion 6}: SpecCoder maintains stronger specification-aware code generation across problem difficulties and provides a more effective backbone for training-free agent-based workflows, suggesting that the proposed specification-aware training strategy benefits both standalone generation and iterative self-repair.
\end{tcolorbox}

\subsection{\textcolor{black}{Specification Alignment and Dependency Analysis}}

\textcolor{black}{
This section systematically examines the role of structured specifications in SpecCoder from multiple perspectives. 
First, we conduct a human evaluation to assess their coverage of the original requirements and their consistency with the generated code. 
Since consistency alone is insufficient to establish that the model actually relies on the specification during inference, we further conduct controlled perturbation experiments to examine the dependence of both code generation and code discrimination on specification semantics. 
% This forms a progressive analysis from representation quality to task-level dependency.
}

\subsubsection{\textcolor{black}{Specification Alignment Evaluation}}

\textcolor{black}{
\textbf{Setup.}
To assess specification quality and its alignment with generated code, we manually evaluate intermediate reasoning on 150 stratified samples from the three benchmarks while preserving their original difficulty distributions.
For each sample, two software engineering researchers independently assess the intermediate reasoning and corresponding code generated by Free-form, Planning-CoT, and SpecCoder, with method labels hidden and outputs randomly ordered. The evaluation focuses on semantic content rather than output format and follows the seven-dimensions. 
\textbf{Requirement Coverage} measures how completely the intermediate reasoning captures the original problem requirements, while \textbf{Spec-Code Consistency} measures how well the generated code aligns with the requirements, logic, and constraints expressed in the intermediate reasoning. Detailed scoring rubrics are provided in~\ref{appendix_spec_eval}.}

\begin{table*}[!t]
  \centering
  \footnotesize 
  % \color{blue} 
  \setlength{\extrarowheight}{3pt} 
  \setlength{\tabcolsep}{8pt} 
  \begin{tabular}{l|c|c}
    \hline
    \textbf{IR} & \textbf{Requirement Coverage} & \textbf{Spec-Code Consistency} \\
    \hline
    Free-form & $3.42 \pm 0.15$ & $3.35 \pm 0.18$ \\
    Planning-CoT & $3.68 \pm 0.14$ & $3.52 \pm 0.16$ \\
    \textbf{SpecCoder (Ours)} & $\mathbf{4.55 \pm 0.12}$ & $\mathbf{4.48 \pm 0.14}$ \\
    \hline
  \end{tabular}
  \caption{\textcolor{black}{Human evaluation results for specification alignment based on a 5-point Likert scale, reported as mean values with 95\% confidence intervals.}}
  \label{tab:rebuttal_human_eval}
\end{table*}

\textcolor{black}{
\textbf{Results and Analysis.}
As shown in Table~\ref{tab:rebuttal_human_eval}, the human evaluation shows substantial inter-rater agreement, with weighted Cohen's $\kappa$ values of 0.81 and 0.77 for Requirement Coverage and Spec-Code Consistency, respectively.
SpecCoder achieves the highest scores of 4.55 and 4.48 on the two metrics, while Free-form and Planning-CoT score between 3.3 and 3.7.
The 95\% confidence intervals further show a clear performance gap.
These results indicate that structured specifications capture the original requirements more completely and exhibit stronger consistency with the generated code.
}

\subsubsection{\textcolor{black}{Dependency in Code Generation}}
\label{sec:dependency_gen_task}

\textcolor{black}{\textbf{Setup.}
To investigate whether code generation depends on the structured specification, we conduct inference-time perturbation experiments across three benchmarks. For each problem, we perturb the generated specification while keeping the original problem, model parameters, prompt template, and decoding configurations unchanged. The perturbed specification is provided within the \texttt{<ANALYZE>} tags as fixed context for generation from the \texttt{<CODE>} tag. The settings are as follows:
}

\textcolor{black}{
\begin{itemize}
    \item \textbf{Empty}: Removes all specification content while retaining the format tags.
    \item \textbf{Removed}: Randomly removes one dimension from the original specification while preserving the remaining dimension.
    \item \textbf{Shuffled}: Replaces the original specification with one from another problem of comparable length.
    \item \textbf{Corrupted}: Alters selected semantic constraints, such as numerical boundaries or I/O requirements, while leaving the remaining content unchanged.
    \item \textbf{Standard}: Uses the original specification without modification.
\end{itemize}
}

\begin{figure}[!t]
    \centering
    \includegraphics[width=0.96\linewidth]{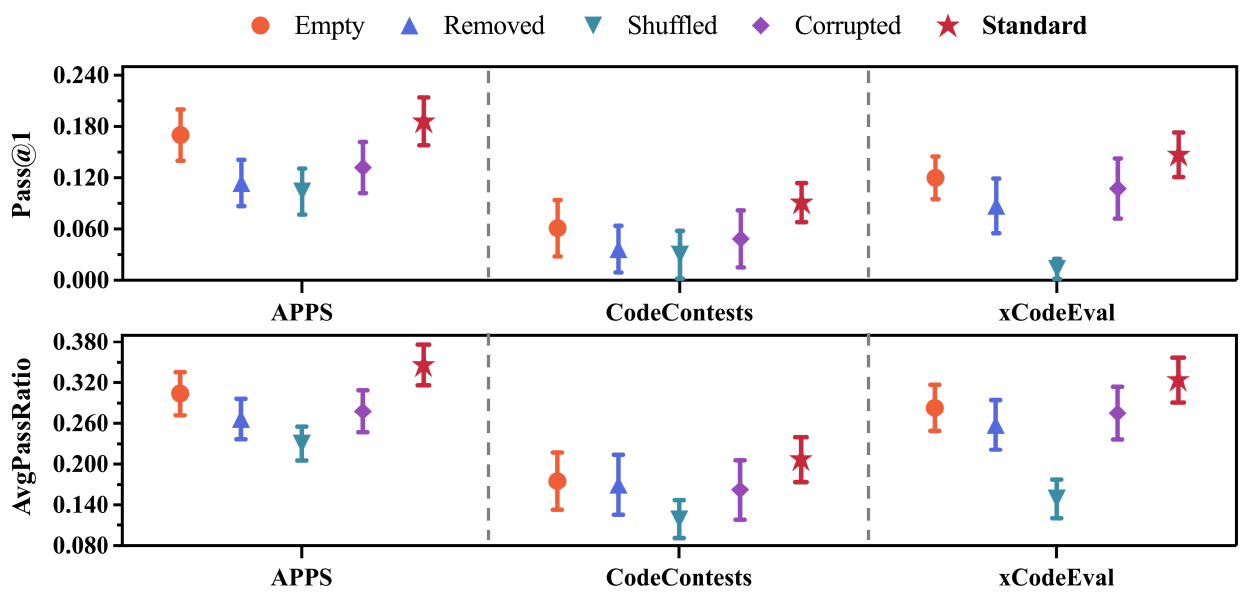}
    \caption{\textcolor{black}{Performance impact of inference-time specification perturbations on code generation using \textit{Qwen2.5-7B-Coder-Instruct}. Pass@1 and AvgPassRatio are reported with 95\% confidence intervals, comparing four perturbation strategies against the Standard baseline.}}
    \label{figs:rebuttal_spec_analysis}
\end{figure}

\textcolor{black}{\textbf{Results and Analysis.}
As shown in Figure~\ref{figs:rebuttal_spec_analysis}, the \textbf{Standard} setting consistently achieves the best performance, while all perturbation settings lead to degradation.
Notably, \textbf{Empty} generally outperforms the other perturbations, indicating that missing specification information is less harmful than providing incomplete or incorrect information.
Among these perturbations, \textbf{Shuffled} causes the largest overall degradation, as illustrated by xCodeEval Pass@1 dropping from 0.147 under \textbf{Standard} to 0.014.
\textbf{Corrupted} also generally performs worse than \textbf{Removed} and falls below \textbf{Empty} in most cases, with APPS Pass@1 decreasing from 0.170 to 0.132.
These results show that the benefit of structured specifications depends critically on their semantic correctness and alignment with the original problem.
This further suggests that SpecCoder has learned to condition code generation on specification semantics, as misleading specifications can be more detrimental than the absence of specification information.
}

\subsubsection{\textcolor{black}{Dependency in Code Discrimination}}

\textcolor{black}{
\textbf{Setup.}
To examine whether code discrimination relies on specification information, we sample 200 test instances and follow the procedure in Section~\ref{sec:code_discrimination_dataset} to construct a fixed candidate pair $(c^+, c^-)$ for each instance, representing correct and defective implementations, respectively.
We then evaluate discrimination accuracy under the same specification perturbations defined in Section~\ref{sec:dependency_gen_task}.
The candidate code pairs are kept identical across all settings, so differences in discrimination accuracy can be primarily attributed to changes in the specification context.
}

\textcolor{black}{
\textbf{Results and Analysis.}
Under the \textbf{Standard} specification, the model achieves 91\% discrimination accuracy, which drops to 68\% when the specification content is removed under \textbf{Empty}.
The corresponding accuracies under \textbf{Removed}, \textbf{Shuffled}, and \textbf{Corrupted} are 60\%, 43\%, and 49\%, respectively.
All three content perturbations fall below \textbf{Empty}, with \textbf{Shuffled} and \textbf{Corrupted} yielding the lowest accuracies, further highlighting the importance of complete and semantically correct specification information for code discrimination.
These results indicate that code discrimination relies on the semantic correspondence between the structured specification and the candidate implementations.
}

\begin{tcolorbox}[
colback=gray!5!white,
colframe=gray!95!black,
fonttitle=\bfseries,
arc=2.8pt,
boxrule=0.9pt,
enhanced,
boxsep=1.5pt,
left=1.5pt, right=1.5pt,
top=1.5pt, bottom=1.5pt,
before upper={\setstretch{0.9}}
]
\textcolor{black}{\textbf{Conclusion 7}: The perturbation experiments provide further evidence that SpecCoder relies on structured specifications during both code generation and discrimination. Performance degradation under removed, shuffled, or corrupted specifications indicates that the model responds to the semantic content of the specification rather than merely its structured format.}
\end{tcolorbox}

\subsection{\textcolor{black}{Evaluation on Realistic Benchmarks}}

\begin{figure*}[!t]
\centering
\includegraphics[width=\linewidth]{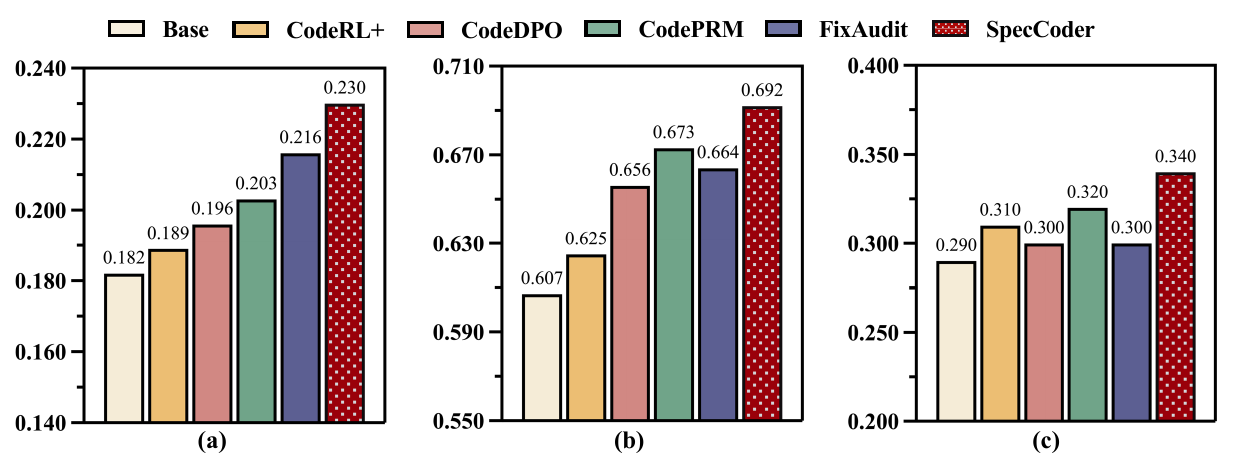}
\caption{\textcolor{black}{Performance comparison in terms of Pass@1 for SpecCoder against the base \textit{Qwen2.5-7B-Coder-Instruct} and four reinforcement learning baselines. (a) BigCodeBench-Hard; (b) ClassEval function-level; (c) ClassEval class-level.}}
\label{figs:rebuttal_real_repo_pass@1}
\vspace{-0.15cm}
\end{figure*}

\textcolor{black}{To further evaluate code generation capabilities in realistic programming scenarios, we conduct additional experiments on BigCodeBench-Hard~\cite{zhuo2025bigcodebench} and ClassEval~\cite{du2023classeval} using the official evaluation scripts. The former focuses on the ability to call complex third-party libraries, while the latter assesses model performance in object-oriented and context-dependent scenarios at both the function and class levels. All results are reported using Pass@1 under greedy decoding.}

\textcolor{black}{As shown in Figure~\ref{figs:rebuttal_real_repo_pass@1}, SpecCoder achieves the best performance across all evaluated settings. On BigCodeBench-Hard, SpecCoder reaches 0.34, compared with 0.29 for the base model and 0.32 for CodePRM. 
On ClassEval function-level tasks, SpecCoder achieves 0.692, outperforming CodePRM at 0.673 and FixAudit at 0.664. On class-level tasks, SpecCoder improves the base model Pass@1 from 0.182 to 0.230. 
These results show that SpecCoder extends its performance gains to more realistic code generation settings involving third-party libraries, object-oriented programming, and class-level implementation.}

\begin{figure}[htpb]
    \centering
    \includegraphics[width=\linewidth]{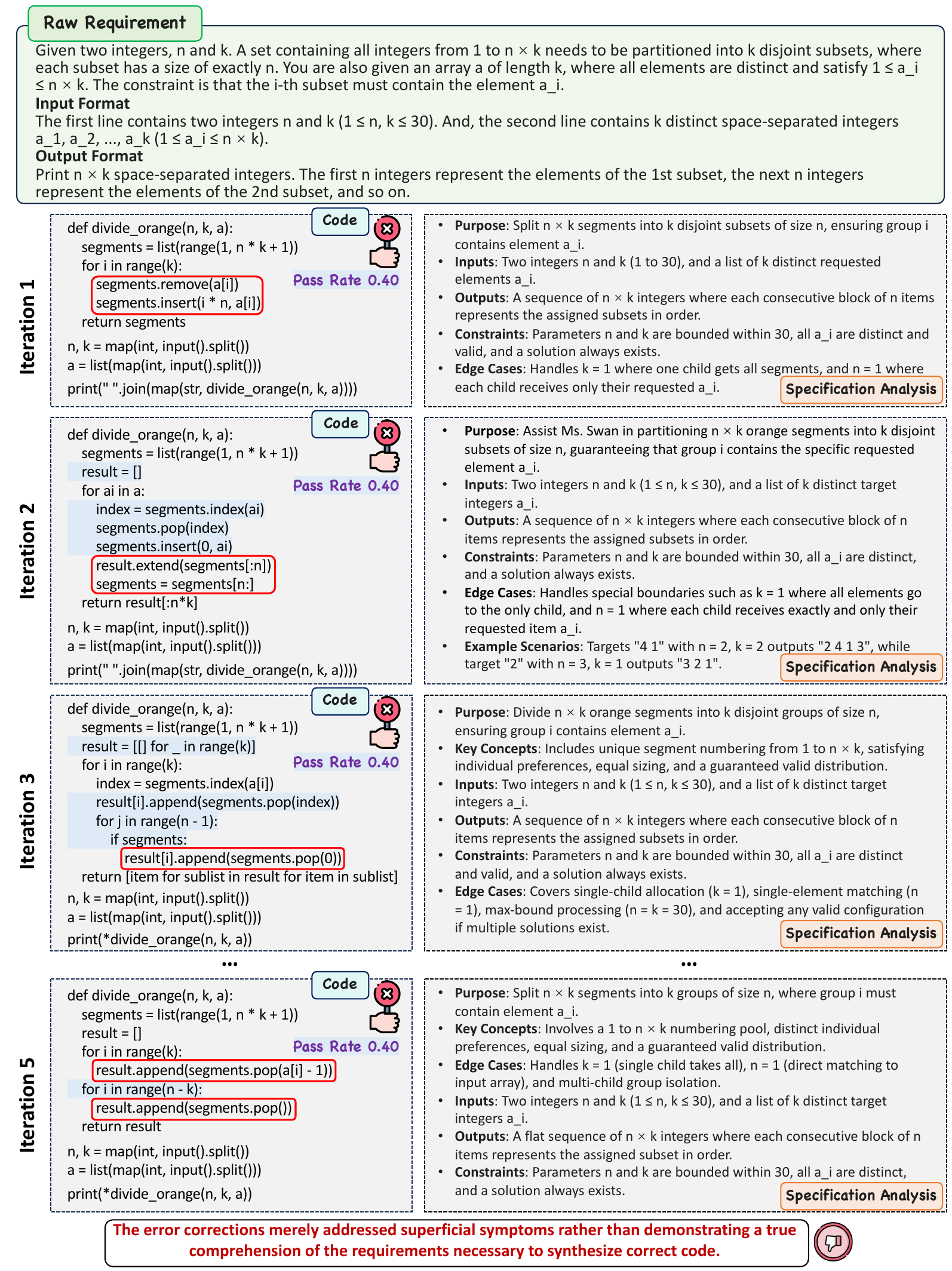}
    \vspace{-0.9cm}
    \caption{Case study of the base \textit{Qwen2.5-7B-Coder-Instruct} model on APPS \texttt{test\_1482}. Within the \textbf{Specine} framework, the figure illustrates the detailed interaction and evolution of code and specifications across five iterations.}
    \label{figs:case_study_base}
\end{figure}

\subsection{Case Studies}

To qualitatively examine specification-aware behavior during iterative code refinement, we select a constrained partition problem from APPS (\texttt{test\_1482}) as the case study.
The task requires partitioning $n \times k$ elements into $k$ disjoint subsets, each of size $n$, where the $i$-th subset must contain the preferred element $a_i$.
This problem is suitable for analysis because correct solutions must preserve strict capacity constraints and element dependencies throughout implementation.

As shown in Figure~\ref{figs:case_study_base}, the base LLM undergoes five refinement iterations, but its test case pass rate remains at $40\%$.
In the first iteration, it removes the preferred elements and inserts them using a static global stride $i \times n$, which causes index drift.
Although later iterations attempt to reconstruct the solution with slicing, the LLM eventually reverts to the earlier flawed framework and changes the insertion position to $i \times n - 1$.
This behavior indicates that the generated specification-level constraints are not consistently reflected in the implementation, leading to a persistent mismatch between requirement understanding and code behavior.

\begin{figure}[!t]
    \centering
    \includegraphics[width=\linewidth]{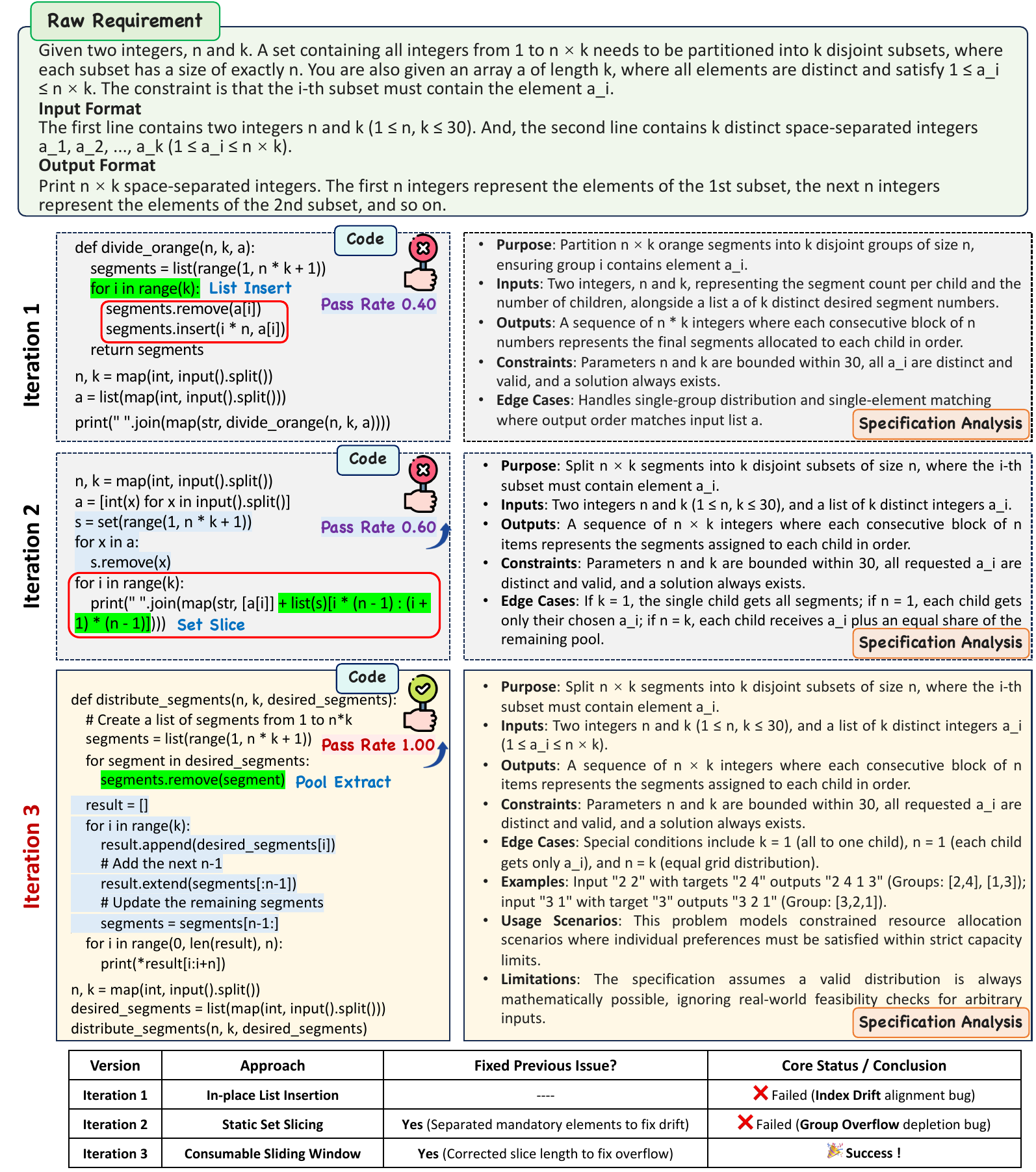}
    \vspace{-0.9cm}
    \caption{Case study of the \textit{Qwen2.5-7B-Coder-Instruct} model trained with \textbf{SpecCoder} on APPS \texttt{test\_1482}. Within the \textbf{Specine} framework, the figure illustrates the interaction and evolution of code and specifications, reaching a 100 percent test case pass rate within three iterations.}
    \label{figs:case_study_grpo}
\end{figure}

In contrast, SpecCoder reaches a test case pass rate of $100\%$ in three iterations, as shown in Figure~\ref{figs:case_study_grpo}. 
It first separates the preferred elements from the remaining pool and then adopts a sliding-window consumption strategy, using \texttt{segments[]} for dynamic padding and truncating the residual pool to avoid index offset errors.
This update preserves both subset capacity and preferred-element constraints, raising the pass rate to $100\%$.
The final iteration further simplifies the code while maintaining the correct partitioning logic.
Meanwhile, the generated structured specification analysis covers relevant edge cases and implementation constraints, providing more stable guidance for refinement.
This case suggests that SpecCoder can better preserve raw requirement constraints during iterative refinement by aligning structured specification analyses with concrete code behavior.
While this qualitative example does not replace systematic quantitative evaluation, it illustrates how specification-aware training can reduce the gap between stated specification-level understanding and generated code behavior.

\section{\textcolor{black}{Discussion}}

\subsection{\textcolor{black}{Cost Analysis}}

\textcolor{black}{
The data construction pipeline contains two API-dependent stages: multi-path generation using \textbf{\textit{GPT-4o}} (\textit{gpt-4o-2024-08-06}) and backward verification using \textbf{\textit{DeepSeek-V3-0324}}~\cite{deepseek2025v30324}, while forward verification relies on local execution and incurs no API fees.
Starting from 20,000 seed problems, we query \textbf{\textit{GPT-4o}} three times per seed, resulting in 60,000 candidate solutions.
Each query contains an average of 880 input tokens and 713 output tokens, corresponding to approximately \$0.028 per seed problem and a total generation cost of \$559.8.
Among these candidates, 88.7\% pass forward verification, leaving 53,220 candidates for backward verification.
Each \textbf{\textit{DeepSeek-V3-0324}} query consumes an average of 1,592 input tokens and 244 output tokens, resulting in approximately \$0.000698 per candidate and \$37.16 in total.
After bidirectional filtering, 13,532 high-quality triplets are retained for $\mathcal{D}_{\mathrm{gen}}$.
The total API expenditure is therefore approximately \$596.96, of which generation and backward verification account for 93.8\% and 6.2\%, respectively.
This corresponds to an amortized API cost of approximately \$0.044 per retained sample, showing that the proposed pipeline maintains a relatively low API cost despite the stringent filtering process.
}

\subsection{\textcolor{black}{Limitations and Future Work}}

\textcolor{black}{Although SpecCoder shows effectiveness in challenging code generation tasks, this study has limitations to address in future work.}
\textcolor{black}{\textbf{First, data construction process relies heavily on external teacher models.}
This ensures high quality training data but introduces additional API costs. 
More importantly, the ability of our model to understand specifications is limited by the capabilities of these teacher models. 
Future work will explore ways to automatically synthesize high quality data without relying on closed source models. 
We plan to achieve this through mechanisms like self-play~\cite{yuan2024self,zhang2025process} and weak to strong generalization.
}
\textcolor{black}{\textbf{Second, executable test cases have limitations as reward signals.} 
During the reinforcement learning stage, we use the pass rate on test cases as an approximate signal to measure whether the code meets the specification. 
However, existing test suites are often incomplete. 
They mainly verify input and output correctness through black box testing. 
They struggle to cover all extreme edge cases and cannot directly check if the code follows specific internal logic or external constraints. 
Future research will leverage static code analysis or process reward models for multidimensional specification feedback beyond basic execution results.
}

\section{Conclusion}

In this paper, we propose SpecCoder, a specification-aware two-stage training framework for improving LLM-based code generation on challenging programming tasks with complex natural language requirements.
SpecCoder introduces structured specification analyses as an intermediate interface between raw requirements and source code, and combines specification-guided SFT with curriculum dual-task GRPO to optimize both code generation and code discrimination.
This design enables LLMs to explicitly derive structured specification analysis from raw requirements, generate code conditioned on them, \textcolor{black}{and strengthen the correspondence between specification-level understanding and generated code behavior.
Experiments on APPS, CodeContests, and xCodeEval show that SpecCoder consistently improves standalone and agent-based code generation, while additional evaluations on BigCodeBench-Hard and ClassEval demonstrate its effectiveness in more realistic programming scenarios.
Ablation studies, human evaluation, and specification perturbation analyses further show that the proposed training components contribute to overall performance and that the model relies on structured specification semantics during code generation and discrimination.}

\clearpage
\appendix
\section{Data Construction Prompts and Verification}
\label{appendix_prompt}

\subsection{Multi-Path Generation and Verification Prompts}

\begin{figure*}[ht]
\centering
\includegraphics[width=0.9\linewidth]{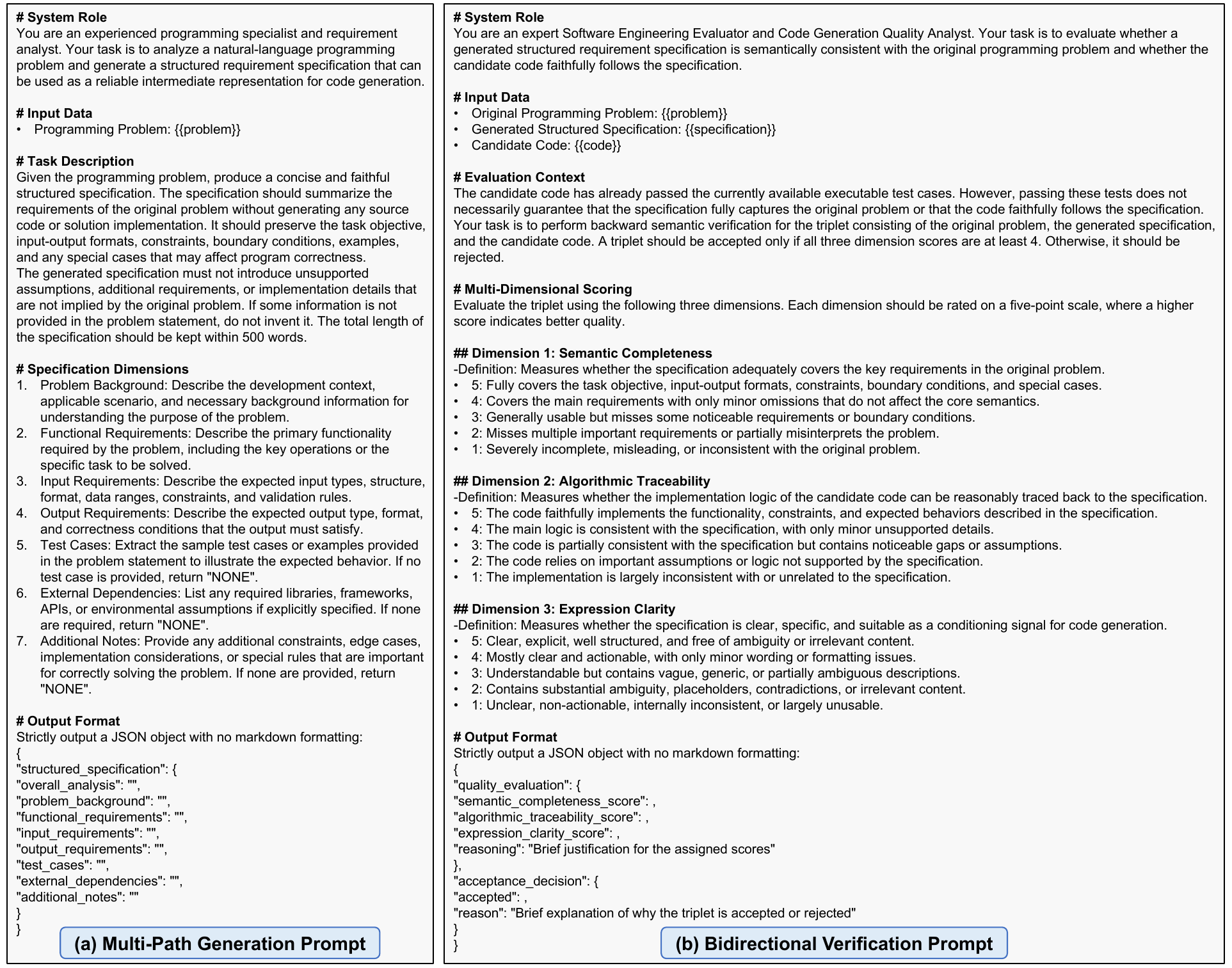}
\caption{Prompt templates used in the SpecCoder data construction pipeline. Subfigure (a) shows the multi-path specification generation prompt, which guides \textbf{\textit{GPT-4o}} (\textit{version: gpt-4o-2024-08-06}) to generate structured specification analyses from raw requirements. Subfigure (b) shows the bidirectional verification prompt, which guides \textbf{\textit{DeepSeek-V3-0324}} to evaluate each triplet along semantic completeness, algorithmic traceability, and expression clarity.}
\label{figs:prompts}
\vspace{-0.15cm}
\end{figure*}

To improve reproducibility, we provide the prompt templates used in the construction of $\mathcal{D}_{\mathrm{gen}}$. As shown in Figure~\ref{figs:prompts} part(a), the multi-path generation prompt instructs \textbf{\textit{GPT-4o}} (\textit{version: gpt-4o-2024-08-06}) to produce structured specification analyses according to the seven predefined dimensions, while the bidirectional verification prompt instructs \textbf{\textit{DeepSeek-V3-0324}} to evaluate the semantic consistency of each triplet $\langle p, s, c\rangle$ before it is retained for training.

\subsection{Bidirectional Verification Evaluation Rubrics and Prompt}
\label{appendix_verification}

This section describes the evaluation rubrics used in the backward verification stage of the data construction pipeline.
After forward execution-based verification, a candidate implementation may pass all available test cases, but its corresponding structured specification analysis may still omit key requirements, introduce unsupported assumptions, or fail to faithfully reflect the implemented behavior.
Therefore, for each triplet $\langle p, s, c\rangle$, where $p$ is the raw requirement, $s$ is the generated structured specification analysis, and $c$ is the candidate implementation, we use \textbf{\textit{DeepSeek-V3-0324}}~\cite{deepseek2025v30324} as an automatic evaluator to assess semantic consistency before adding the triplet to $\mathcal{D}_{\mathrm{gen}}$.
Backward verification is conducted along three dimensions:

\begin{itemize}
\item \textbf{Semantic Completeness.}
This dimension evaluates whether the structured specification analysis $s$ preserves the key requirement information in the raw requirement $p$, including the task objective, input and output formats, constraints, boundary conditions, and special cases.

\item \textbf{Algorithmic Traceability.}
This dimension evaluates whether the implementation logic of the candidate code $c$ can be reasonably traced back to the structured specification analysis $s$, including the core algorithm, input/output handling, state updates, and boundary-case processing.

\item \textbf{Expression Clarity.}
This dimension evaluates whether the structured specification analysis $s$ is clear, specific, and suitable as a conditioning signal for code generation, without vague placeholders, ambiguous descriptions, contradictions, or irrelevant content.
\end{itemize}

Each dimension is rated on a five-point scale, as detailed in Figure~\ref{figs:prompts} (b).
A triplet is retained in $\mathcal{D}_{\mathrm{gen}}$ only if it receives a score of at least 4 in all three dimensions.
Unlike forward verification, which focuses on executable correctness, backward verification provides a fine-grained assessment of the semantic consistency among the raw requirement, structured specification analysis, and generated implementation.
This process filters out triplets that pass the available tests but contain incomplete, inconsistent, or unclear specification analyses.

To assess the reliability of the automatic evaluation, we conduct a manual audit of 400 triplets randomly sampled from the 13,532 triplets accepted by the automatic evaluator.
According to Cochran's formula~\cite{cochran1977sampling}, this sample size corresponds to a 95\% confidence level with an approximate margin of error of 4.83\% under the conservative maximum-variance assumption.
Two researchers with expertise in software engineering and program analysis independently evaluate the sampled triplets using the same three dimensions and five-point scoring criteria.
Both annotators are blind to the scores assigned by the automatic evaluator.
Disagreements are resolved through discussion, with a third annotator consulted when necessary.

The two annotators achieve a Cohen's Kappa of $\kappa=0.79$, indicating substantial inter-annotator agreement~\cite{landis1977measurement}.
After resolving disagreements, 88\% of the automatically accepted triplets also satisfy the human acceptance criterion, requiring a score of at least 4 in all three dimensions.
This human-validated acceptance rate provides empirical support for using automatic backward verification to construct high-quality specification-guided generation data at scale.

\section{Expanded Descriptions of Benchmark Datasets}
\label{appendix_data_info}

To evaluate SpecCoder, we adopt three competitive code generation benchmarks: APPS~\cite{hendrycks2021measuring}, CodeContests-raw~\cite{li2022competition}, and xCodeEval~\cite{khan2024xcodeeval}. To construct the required specification-based generation and discrimination tasks, the data is processed via bidirectional filtering. The subsequent subsections provide detailed descriptions and difficulty classifications for each dataset.

\subsection{APPS}
This dataset collects programming problems from various online platforms such as Codeforces and LeetCode. It contains 5,000 training samples and 5,000 test samples. 
For convenience, we denote the three difficulty levels of introductory, interview, and competition as easy, medium, and hard respectively.
In the training phase, we utilize all training data. 
To obtain richer training resources, we also select 3,000 samples from the test set for data generation according to the original difficulty distribution. 
In the testing phase, following previous work~\cite{tian2025ufiX,li2026bridging}, we sample 500 problems from the test set for our experiments to balance evaluation cost and statistical representation. 
This sub-sample strictly maintains the original difficulty distribution across easy, medium, and hard levels.

\subsection{CodeContests}
It is introduced by Google DeepMind to evaluate the ability of models to solve competitive programming problems. The dataset consists of 13,610 training problems and 165 test problems.
Based on the platform rating score $R$, the problems are divided into three levels, where easy represents $800 \le R \le 1400$, medium represents $1400 < R \le 1900$, and hard represents $R > 1900$.
In the training phase, we utilize all 13,610 training problems as base data for data construction during the generation stage. In the testing phase, we use the complete test sets for evaluation.

\subsection{xCodeEval}
It is a large-scale competitive code generation benchmark that contains about 7,500 programming problems. 
The problems are divided into three levels based on the difficulty value $D$, where easy represents $D \le 1400$, medium represents $1400 < D \le 2000$, and hard represents $D > 2000$.
In the training phase, this dataset does not participate in the construction of the training data to ensure a fair evaluation. In the testing phase, we use it purely as a test set to check the robustness of the models. 
Following the settings of previous work~\cite{tian2025ufiX,li2026bridging}, we sample a subset of 300 problems from this benchmark to evaluate our method and the baseline models.

\section{Human Evaluation Rubrics}
\label{appendix_spec_eval}

\begin{figure*}[ht]
\centering
\includegraphics[width=\linewidth]{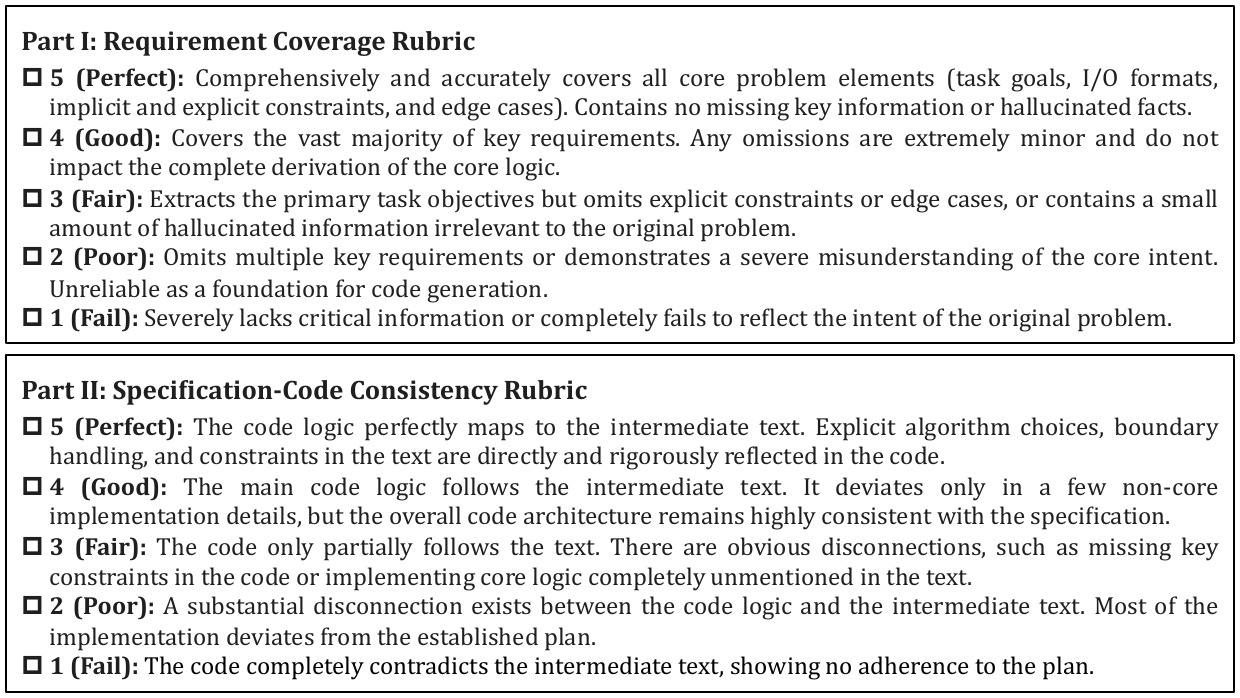}
\caption{\textcolor{black}{The 5-point Likert rubric for evaluating specification alignment.}}
\label{figs:rebuttal_human_eval}
\vspace{-0.15cm}
\end{figure*}

\textcolor{black}{To ensure a fair and consistent assessment across different intermediate representations, the human evaluation employs a 5-point Likert scale. 
The seven predefined specification dimensions in Table~\ref{tab:spec_dimensions} are used only as a semantic reference rather than as separately scored items or required output fields. 
Evaluators examine whether the intermediate reasoning captures the requirement information represented by these dimensions, regardless of how such information is organized or expressed. 
Based on this assessment, Requirement Coverage measures the overall completeness of the captured requirements, while Spec-Code Consistency evaluates how consistently the generated code follows the requirements, logic, and constraints expressed in the intermediate reasoning. Detailed scoring criteria are provided in Figure~\ref{figs:rebuttal_human_eval}.}

\begin{table}[htbp]
    \centering
    \small
    \begin{tabular}{l c || l c}
    \hline
    \multicolumn{4}{c}{\textbf{Specification-Guide SFT Stage} (Model: Qwen2.5-7B-Coder-Instruct-base)} \\ \hline
    \textbf{Hyperparameter} & \textbf{Value} & \textbf{Hyperparameter} & \textbf{Value} \\ \hline
    LR Scheduler & Cosine & Cutoff Length & 4096 \\
    Learning Rate & $1.0 \times 10^{-4}$ & Precision & BF16 \\
    Epochs & 3.0 & LoRA Rank & 8 \\
    Warmup Ratio & 0.1 & LoRA Target & all modules \\ \hline
    \multicolumn{4}{c}{\textbf{Curriculum Dual-task GRPO Stage} (Model: Qwen2.5-7B-Coder-Instruct-SFT)} \\ \hline
    \textbf{Hyperparameter} & \textbf{Value} & \textbf{Hyperparameter} & \textbf{Value} \\ \hline
    Parallel Strategy & FSDP2 & Max Prompt Length & 2048 \\
    Policy Learning Rate & $1.0 \times 10^{-6}$ & Max Generate Length & 1024 \\
    Samples per Prompt & 8 & Inference Engine & vLLM \\
    Train Batch Size & 16 & Epochs & 3 \\ \hline
    \end{tabular}
    \caption{Hyperparameter settings for the two-stage training of SpecCoder.}
    \label{tab:hyperparameters}
\end{table}

\begin{table}[htbp]
\centering
\small
% \color{blue}
\begin{tabular}{c|cc|ccc}
\hline
\multirow{2}{*}{\textbf{Difficulty}} & \multicolumn{2}{c|}{\textbf{Task Types}} & \multicolumn{3}{c}{\textbf{Curriculum Dual-task GRPO Stage}} \\
\cline{2-6}
 & \textbf{Generation} & \textbf{Discrimination} & \textbf{Phase 1} & \textbf{Phase 2} & \textbf{Phase 3} \\
\hline
Easy   & 2,408 & 1,695 & \checkmark\checkmark &                      & \checkmark \\
Medium & 2,701 & 1,004 &                      & \checkmark\checkmark & \checkmark \\
Hard   & 1,702 & 710   &                      & \checkmark\checkmark & \checkmark \\
\hline
\end{tabular}
\caption{\textcolor{black}{Data distribution and epoch allocation across the three GRPO phases. Each checkmark (\checkmark) represents one training epoch for the corresponding subset.}}
\label{tab:data_distribution}
\end{table}

\section{Implementation and Hyperparameter Configurations}
\label{appendix_hyper}

\textcolor{black}{We conduct the two-stage training of SpecCoder on two NVIDIA A100-SXM4 GPUs using a random seed of 42 for training and evaluation. 
The corresponding hyperparameter configurations are provided in Table~\ref{tab:hyperparameters}. 
The supervised fine-tuning stage uses 13,532 samples and takes 1.50 hours, with a peak memory usage of 73,058 MB per GPU.
The curriculum dual-task GRPO stage uses 10,220 samples, including 6,811 generation samples and 3,409 discrimination samples, which are further partitioned into easy, medium, and hard subsets. 
Within each curriculum stage, generation and discrimination samples from the designated difficulty subsets are randomly interleaved, with a peak memory usage of 76,445 MB per GPU. 
As illustrated in Figure~\ref{figs:overview} and Table~\ref{tab:data_distribution}, the three curriculum stages require 13.17 hours for easy warmup, 25.67 hours for hard focus, and 18.38 hours for full consolidation. 
Each stage initializes the model from the checkpoint of the preceding stage while resetting the optimizer state. As formalized in Algorithm~\ref{alg:training}, easy samples are processed twice in the first stage and once in the third, whereas medium and hard samples are processed twice in the second stage and once in the third. 
Thus, each sample is seen exactly three times, matching the total sample exposure of a standard three-epoch training scheme.
}

\begin{algorithm}[htbp]
\caption{SpecCoder Two-Stage Training}
\label{alg:training}
% \color{blue}
\SetAlFnt{\footnotesize} 
\renewcommand{\baselinestretch}{0.85}\selectfont
\setlength{\algomargin}{0.045\linewidth}
\SetKwInput{KwIn}{Input}
\SetKwInput{KwOut}{Output}

\textbf{Specification-Guide SFT Stage}\\
\KwIn{Pre-trained policy $\pi_{\theta_{\mathrm{init}}}$; Generation dataset $\mathcal{D}_{\mathrm{gen}}$}

$\theta \gets \theta_{\mathrm{init}}$\;
\For{epoch $= 1$ \KwTo $E_{\mathrm{SFT}}$}{
    $\mathcal{B}_{\mathrm{SFT}} \gets \mathrm{Shuffle}(\mathcal{D}_{\mathrm{gen}})$\;
    \For{mini-batch $B \subset \mathcal{B}_{\mathrm{SFT}}$}{
        $\theta \gets \theta - \eta \nabla \mathcal{L}_{\mathrm{SFT}}$ 
    }
}
$\theta_{\mathrm{SFT}} \gets \theta$ 

\vspace{1mm}\hrule\vspace{1mm}

\textbf{Curriculum Dual-task GRPO Stage}\\
\KwIn{SFT policy $\pi_{\theta_{\mathrm{SFT}}}$; Subsets $\{(S_k, E_k)\}_{k=1}^{3}$ containing easy, medium, and hard partitions of $\mathcal{D}_{\mathrm{gen}} \cup \mathcal{D}_{\mathrm{disc}}$}

$\theta_0 \gets \theta_{\mathrm{SFT}}$\;

\For{phase $k = 1, 2, 3$}{
    $\theta \gets \theta_{k-1}$ 
    Reset optimizer state\;
    $\mathcal{B}_k \gets \mathrm{Shuffle}(S_k)$
    
    \For{epoch $= 1$ \KwTo $E_k$}{
        \For{mini-batch $B \subset \mathcal{B}_k$}{
            \For{prompt $x_i \in B$}{
                Sample $G$ responses $\{y_i^{(g)}\}_{g=1}^{G} \sim \pi_\theta(\cdot \mid x_i)$\;
                Compute rewards $\{r_i^{(g)}\}_{g=1}^{G}$ via execution or spec matching\;
                Compute group-relative advantages $\hat{A}_i^{(g)}$\;
            }
            $\theta \gets \theta + \eta \nabla \mathcal{L}_{\mathrm{GRPO}}$ 
        }
    }
    $\theta_k \gets \theta$
}

\vspace{1mm}\hrule\vspace{1mm}
\KwOut{Final optimized policy $\pi_{\theta_3}$}
\end{algorithm}

\section{Reproducibility and Artifact Availability}
\label{appendix_code}
\textcolor{black}{We use a random seed of 42 for the main training and evaluation experiments. The exact training and evaluation samples, together with their task identifiers, are publicly released to reproduce the data splits used in our experiments.
The code and more information are publicly available at \url{https://github.com/yixuanli1230/SpecCoder}.}

%% bibitems, please  use




% %% For numbered reference style

%   \textit{\LaTeX: a document preparation system},
%   Addison Wesley, Massachusetts,

% \end{thebibliography}

\clearpage
\bibliographystyle{elsarticle-num}
\bibliography{ref}

\end{document}